\documentclass[journal]{IEEEtran}
\usepackage{cite}
\usepackage{amssymb}
\usepackage{amsbsy}
\usepackage{amsmath,bm}
\usepackage{amsfonts}
\usepackage{array}
\usepackage{graphicx}
\usepackage{multirow}
\usepackage{stfloats}
\usepackage{url}
\usepackage{verbatim}
\usepackage{caption}
\usepackage{subfigure}
\usepackage{svg}
\usepackage{bookmark}
\usepackage{textcomp}
\usepackage{academicons}
\definecolor{orcidlogocol}{HTML}{A6CE39}
\usepackage{threeparttable}
\usepackage{xfrac}
\usepackage{framed}
\colorlet{shadecolor}{yellow!40}
\usepackage{soul}
\usepackage{tabularx}
\usepackage{nopageno}
\usepackage{algorithm,algpseudocode}
\usepackage{mathtools}
\usepackage{latexsym}
\usepackage{tikz}

\usepackage{tikz,xcolor}
\hypersetup{hidelinks}

\definecolor{lime}{HTML}{A6CE39}
\DeclareRobustCommand{\orcidicon}{%
	\begin{tikzpicture}
	\draw[lime, fill=lime] (0,0) 
	circle [radius=0.16] 
	node[white] {{\fontfamily{qag}\selectfont \tiny ID}};
	\draw[white, fill=white] (-0.0625,0.095) 
	circle [radius=0.007];
	\end{tikzpicture}
	\hspace{-2mm}
}

\foreach \x in {A, ..., Z}{%
	\expandafter\xdef\csname orcid\x\endcsname{\noexpand\href{https://orcid.org/\csname orcidauthor\x\endcsname}{\noexpand\orcidicon}}
}

\def\BibTeX{{\rm B\kern-.05em{\sc i\kern-.025em b}\kern-.08em
   T\kern-.1667em\lower.7ex\hbox{E}\kern-.125emX}}

\begin{document}


\title{Advanced Maximum Adhesion Tracking Strategies in Railway Traction Drives\\ 

\thanks{Ahmed Fathy Abouzeid, Juan Manuel Guerrero and Fernando Briz are with the Department of Electrical Engineering, University of Oviedo, 33204 Gij\'{o}n, Spain (e-mail: abouzeidahmed@uniovi.es; guerrero@uniovi.es; fbriz@uniovi.es). Ahmed Fathy Abouzeid is on leave with the Department of Electrical Engineering, Port-Said University, Egypt (e-mail: ahmed\_abouzeid@eng.psu.edu.eg).}%
\thanks{Lander Lejarza, Iker~Muniategui, Aitor~Endema\~{n}o are with the Department of Traction systems, Ingeteam Power Technology, Zamudio, 48170 Spain (e-mail: lander.lejarza@ingeteam.com; iker.muniategui@ingeteam.com; aitor.endemano@ingeteam.com).}%
\thanks{(Corresponding author: Ahmed Fathy Abouzeid).}
\thanks{This work was supported in part by the Government of the Principality of Asturias under Project AYUD/2021/50988. The work of Ahmed Fathy Abouzeid was supported in part by the Scholarship from the Ministry of Higher Education and Scientific Research of Egypt.}
}


\author{Ahmed Fathy Abouzeid,~\IEEEmembership{Member,~IEEE}, Juan Manuel Guerrero,~\IEEEmembership{Senior Member,~IEEE}, Lander Lejarza, Iker Muniategui, Aitor Endemaño, Fernando Briz,~\IEEEmembership{Senior Member,~IEEE}} 


\maketitle
	\thispagestyle{empty}

\begin{abstract}
Modern railway traction systems are often equipped with anti-slip control strategies to comply with performance and safety requirements. A certain amount of slip is needed to increase the torque transferred by the traction motors onto the rail. Commonly, constant slip control is used to limit the slip velocity between the wheel and rail avoiding excessive slippage and vehicle derailment. This is at the price of not fully utilizing the train's traction and braking capabilities.  Finding the slip at which maximum traction force occurs is challenging due to the non-linear relationship between slip and wheel-rail adhesion coefficient, as well as to its dependence on rail and wheel conditions. Perturb and observe (P\&O) and steepest gradient (SG) methods have been reported for the Maximum Adhesion Tracking (MAT) search. However, both methods exhibit weaknesses. 
Two new MAT strategies are proposed in this paper which overcome the limitations of existing methods, using Fuzzy Logic Controller (FLC) and Particle Swarm Optimization (PSO) respectively. 
Existing and proposed methods are first simulated and further validated experimentally using a scaled roller rig under identical conditions. The results show that the proposed methods improve the traction capability with lower searching time and oscillations compared to existing solutions. Tuning complexity and computational requirements will also be shown to be favorable to the proposed methods.
\end{abstract}
\section{Introduction}  \label{sec:Intro}

\IEEEPARstart{T}{ransportation} electrification has become more dominant recently in the transport sector for reducing greenhouse gas emissions and mitigating the effects of climate change on the planet \cite{georgatzi2020examining, mccollum2014transport}. Electric railways, amongst other means of transportation, offer substantially better energy efficiency, lower emissions, and lower operating costs \cite{lin2021impact}. Moreover, electric trains have a superior power-to-weight ratio compared to trains powered by onboard fuel tanks. This allows faster acceleration, higher power, and speed limits with less noise pollution production. Unfortunately, railway electrification capital cost is the main disadvantage as it requires new infrastructure including power supply stations, overhead lines, signaling systems, interference protection circuits, etc \cite{chen2016impact}. Therefore, optimized solutions for railway electrification should be considered during the design and operation phases to achieve the desired revenue \cite{ronanki2017comprehensive, krastev2016future}. 

Optimized traction/braking force utilization is a key aspect of modern railway traction systems for multiple reasons, including safety, performance, reliability, and energy efficiency. Traction force is defined as the force developed by the traction motor being transferred to the rail through the train vehicle's wheel to achieve the desired train speed. Maximizing the traction force leads to more efficient and faster acceleration/deceleration rates. This allows for achieving the planned travel speed-distance profile precisely, avoiding trip delays and reducing energy consumption. Therefore, the implementation of control methods able to maximize the traction force becomes crucial to traction system manufacturers and train service providers.

The traction force that can be transferred to the rail will be limited by the friction between the driven wheels and the steel rail. The adhesion limit will depend on the normal load and the friction coefficient of the contact point, also known as the adhesion coefficient. Adhesion coefficient is a non-linear function slip \cite{polach2005creep}. The slip/skid phenomenon occurs when the traction force surpasses the adhesion limit during traction/braking. Excessive slip or skid will result in an increase in wheel wear and a reduction of the overall traction performance. Therefore, many efforts have been devoted to limiting the slip/skid between the wheel and the rail \cite{kondo2012anti, pichlik2014overview, uyulan2018comparison}. Both direct and indirect re-adhesion control methods have been proposed to limit the slip within a predefined threshold \cite{park2001hybrid, watanabe2000anti, ryoo2003novel, yamashita2015anti}. The main demerit of these traditional re-adhesion methods is that the traction capability is not fully utilized. 

Finding the slip velocity at which the maximum adhesion occurs is a challenging task. This is due to the unpredictability of the wheel-rail contact condition, and consequently the difficulty of estimating the adhesion coefficient. In \cite{schwartz1992regelung}, Perturb and Observe (P\&O) method similar to those used for photovoltaic panels was applied for MAT. The slip controller increases the slip velocity command gradually and the tractive force is monitored and stored. If the maximum point is overstepped, the slip velocity command is decreased to bring the operating point back to the stable region. In \cite{kawamura2002maximum} authors propose to use the adhesion force derivative. The slip velocity command is corrected according to the slope of the adhesion curve, i.e., the slip command increases when the slope of the curve is positive which means that the operating point is in the stable region. If the slope of the adhesion curve is negative, the operating point is in the unstable region and the slip command should be decreased. The previous two methods will be discussed in more detail and tested in this paper.
 
 Several approaches have been proposed using Kalman Filters to avoid measurement noise and the computation of derivatives used in \cite{kawamura2002maximum}. However, these methods rely on the mathematical model of the mechanical drive train and require accurate parameter estimates \cite{liu2017novel, pichlik2018locomotive, pichlik2021adhesion}. The authors in \cite{zirek2018adaptive} proposed an adaptive sliding mode control to stabilize wheel slip and improve traction performance but this method requires accurate measurement of the adhesion force, which is not easy to achieve in practice. Additionally, advanced slip control techniques using model predictive control and adhesion swarm intelligence can be found in \cite{sadr2016predictive, wen2019anti, huang2020online, molavi2022robust, zirek2020novel}. Though these methods show good adhesion performance, they either lack experimental validation \cite{wen2019anti, huang2020online, molavi2022robust} or suffer from implementation complexity and high computational burden \cite{sadr2016predictive, zirek2020novel}.

Despite all previous advancements, there is a lack of literature dealing with the experimental validation of maximum traction force tracking techniques and their performance evaluation under similar operating conditions. In addition, the classical methods possess slow dynamic response for peak searching capability and high steady-state error, which increase the torque ripples on the traction machine. Therefore, it is desirable to develop new MAT techniques capable of enhancing the peak tracking capability and mitigating the undesired stresses on the traction machine during the searching process.  

The contributions of this work can be summarized as follows:
\begin{enumerate}
    \item An overview of existing anti-slip control techniques intended for maximum utilization of the available adhesion. 
    \item A new technique based on knowledge-based Fuzzy Logic Control (FLC) is proposed to enhance the dynamic performance of the classical peak search methods. 
    \item A new technique based on Particle Swarm Optimization (PSO) for MAT is proposed and successfully implemented for railways.   
    \item A simulation-based comparative study of existing and proposed methods using MATLAB/Simulink. 
    \item Development and validation of a new simplified design of a scaled roller rig for emulating the wheel-rail contact dynamics, reproducing the slip phenomenon, and suitable therefore for the comparative analysis of anti-slip control techniques in railways.

\end{enumerate}
 
 The structure of the article is summarized as follows: section \ref{sec:system_description} includes an overview of the design and the overall control scheme of the scaled roller rig with wheel-rail contact emulator; section \ref{sec:overview_slip_control} summarizes the existing slip velocity control methods considered for MAT applied in real trains; section \ref{sec:proposed} includes the two proposed MAT strategies; section \ref{sec:Exp} includes an assessment and comparative analysis of all the methods being considered; conclusions are finally discussed in section \ref{sec:conclusions}.   

\section{System Model and Test bench Description}   \label{sec:system_description}

One of the limitations perceived during the review of the state of the art was that the different methods were validated using different platforms. This makes it extremely difficult to perform a fair comparative analysis. To overcome this limitation, all the methods considered in this paper will be tested in the same test rig. System model as well as test rig design and control are presented in this section.
\subsection{Proposed wheel-rail contact emulator}  \label{subsec:scaled_roller_rig}

Roller rigs combined with simulation verification are convenient replicas for evaluating control techniques of railway traction drives during pre-service commissioning. Full-scale or scaled roller rigs can be used for the purpose of producing the same dynamics of the actual train moving on a rail. Scaled roller rigs are preferred due to cost, size, and manufacturing obstacles. However, special care must be taken for the selection of the scaled parameters during the design process to reproduce as precisely as possible the behavior of the full-scale system. In this paper, Manchester Metropolitan University (MMU) method is used \cite{jaschinski1991application}. In this approach, the locomotive mass is considered to be equally distributed amongst the wheels and the nominal linear wheel speed is scaled by $1/5$ factor \cite{jaschinski1999application}. The full design process and scaling parameters of the scaled roller rig used in this paper can be found in \cite{nihal2022design}.

\begin{figure}[tb]
    \centering
    \includegraphics[width=\columnwidth, trim={0cm 2cm 6cm 2cm},clip]{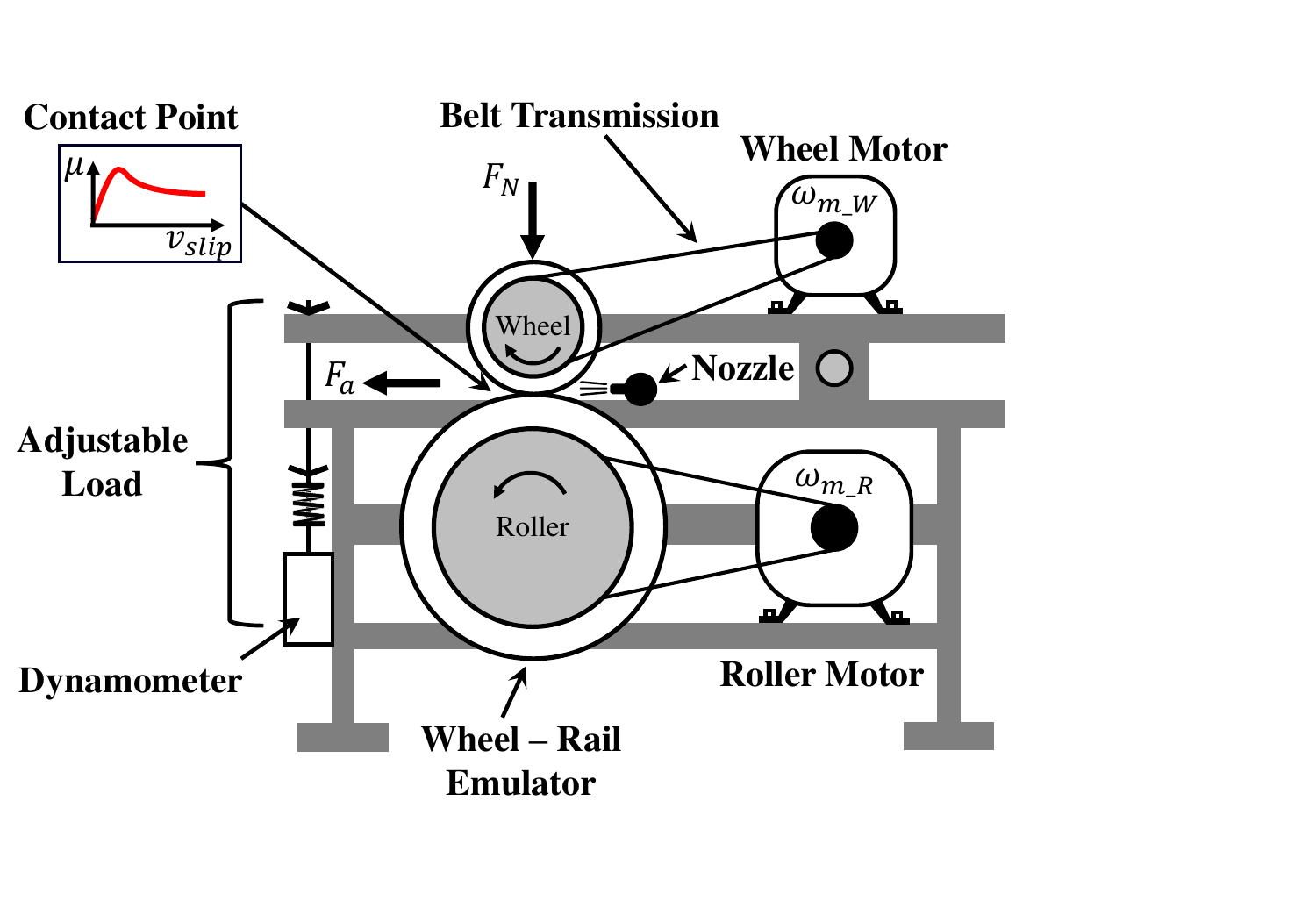}
    \caption{Schematic representation of the proposed scaled roller rig.}
    \label{fig:rollerrig_scheme}
\end{figure}

Fig. \ref{fig:rollerrig_scheme} shows a schematic representation of the scaled roller rig where the smaller wheel represents one of the locomotive wheels and the bigger wheel, referred to as the roller, represents the rail. Two induction motors (IMs) are used to drive both wheels via a transmission belt system. The normal force $F^{}_N$ applied to the wheel is adjusted manually with a spring system including a dynamometer for fine force tuning. The test bench has a water-spraying nozzle to emulate wheel-rail wet conditions.  

Adhesion coefficient $\mu$ is defined in \eqref{eq:mu} as the ratio between the adhesion force $F^{}_a$ being transferred from the wheel to the roller and the normal force applied to the wheel $F^{}_N$. Adhesion is a non-linear function that depends on the slip velocity between the wheel and roller, ambient factors such as humidity and ambient temperature, and the surface condition of the contact point \cite{vollebregt2021challenges}. The slip velocity is given by \eqref{eq:vs}, where $\omega_{m_W}$, $\omega_{m_R}$ and $r^{}_W$, $r^{}_R$ are the mechanical speed of the induction motors (IM) in electrical units (rad/s) and the radius of both wheel and roller respectively.

\begin{align} 
        \mu &= \frac{F^{}_a}{F^{}_N} \label{eq:mu} \\
        v^{}_{slip} &= v^{}_{W} -v^{}_{R}  \label{eq:vs} 
\end{align}

\noindent where $\;\; v^{}_W = \omega_{m_W} \cdot r^{}_W; \;\;  v^{}_R = \omega_{m_R} \cdot r^{}_R $ \\

The adhesion torque (i.e., load torque) can be expressed as the adhesion force exerted on the wheel multiplied by its radius \eqref{eq:Tw}. 

\begin{equation} 
        T^{}_W = F^{}_a \cdot r^{}_W = \mu \cdot F^{}_N \cdot r^{}_W  \label{eq:Tw} 
 \end{equation}

The electromagnetic torque developed by the wheel motor $T_{e\_W}$ and transferred to the wheel $T_W$ via belt transmission is given by \eqref{eq:Tew} where $R_{g\_W}$ is the wheel gear ratio. 

\begin{equation} 
        T_{e\_W} = \frac{T^{}_W}{R_{g\_W}} =  \frac{\mu \cdot F^{}_N \cdot r^{}_W}{R_{g\_W}} \label{eq:Tew} 
 \end{equation}

The same relation can be developed for the roller side where $T_{e\_R}$, and $R_{g\_R}$ are the roller motor electromagnetic torque, and roller gear ratio respectively.  

\begin{equation} 
        T_{e\_R} = \frac{T^{}_R}{R_{g\_R}} =  \frac{\mu \cdot F^{}_N \cdot r^{}_R}{R_{g\_R}} \label{eq:TeR} 
 \end{equation}

\begin{figure*}[t]
    \centering
    \includegraphics[width=0.75\textwidth]{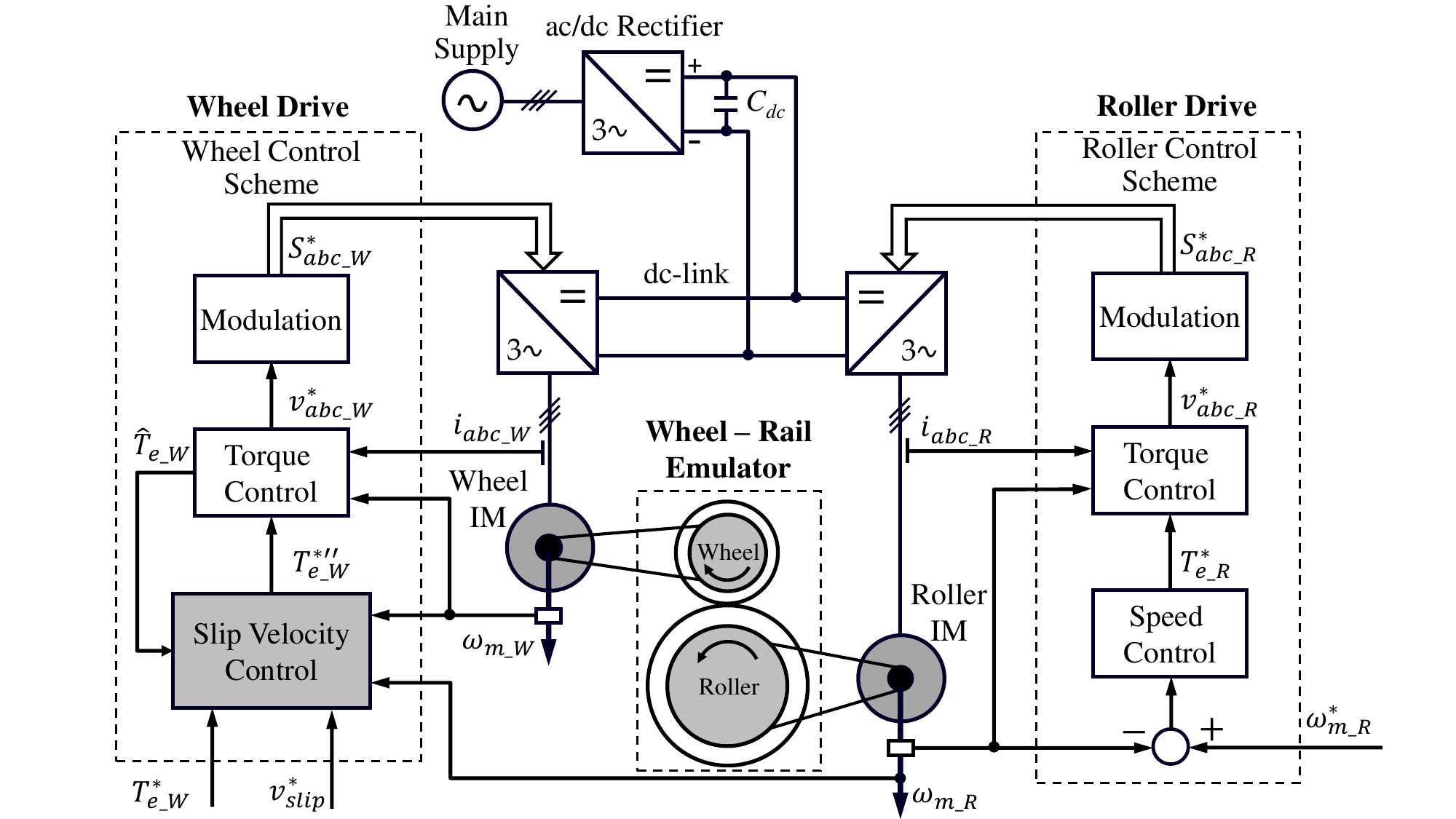}
    \caption{Overall control scheme of proposed scaled roller rig for slip velocity control.}
    \label{fig:general_scheme}
\end{figure*}
\subsection{Proposed scaled roller rig overall control scheme}  \label{subsec:overall_control}

Wheel and roller motors are fed from three-phase inverters which share the same dc link (see Fig. \ref{fig:general_scheme}). The dc link is fed from the grid by means of a diode rectifier. A commercial drive using Rotor Field Oriented Control (RFOC) with an outer speed control loop (see right side of Fig. \ref{fig:general_scheme}) is used to control the roller. 

On the other side, the wheel motor is fed from a custom drive built to implement the different control strategies for MAT functionality. The control scheme of the custom drive contains the same structure as the roller commercial drive except for the speed control block that is replaced by the slip velocity control block (see the gray block on the left side of Fig. \ref{fig:general_scheme}). Further details are provided in Section \ref{sec:Exp}. The slip velocity control block requires additional signals: reference torque $T^{\ast}_{e\_W}$, reference slip velocity $v^{\ast}_{slip}$, and torque estimate $\hat{T}_{e\_W}$. The signals involved depend on the MAT method being used. In all cases, knowledge of roller speed is essential, which implies that real train velocity should be known in the real system. 

\begin{figure}[b]
    \centering
    \includegraphics[width=\columnwidth, trim={0cm 2cm 0cm 3cm},clip]{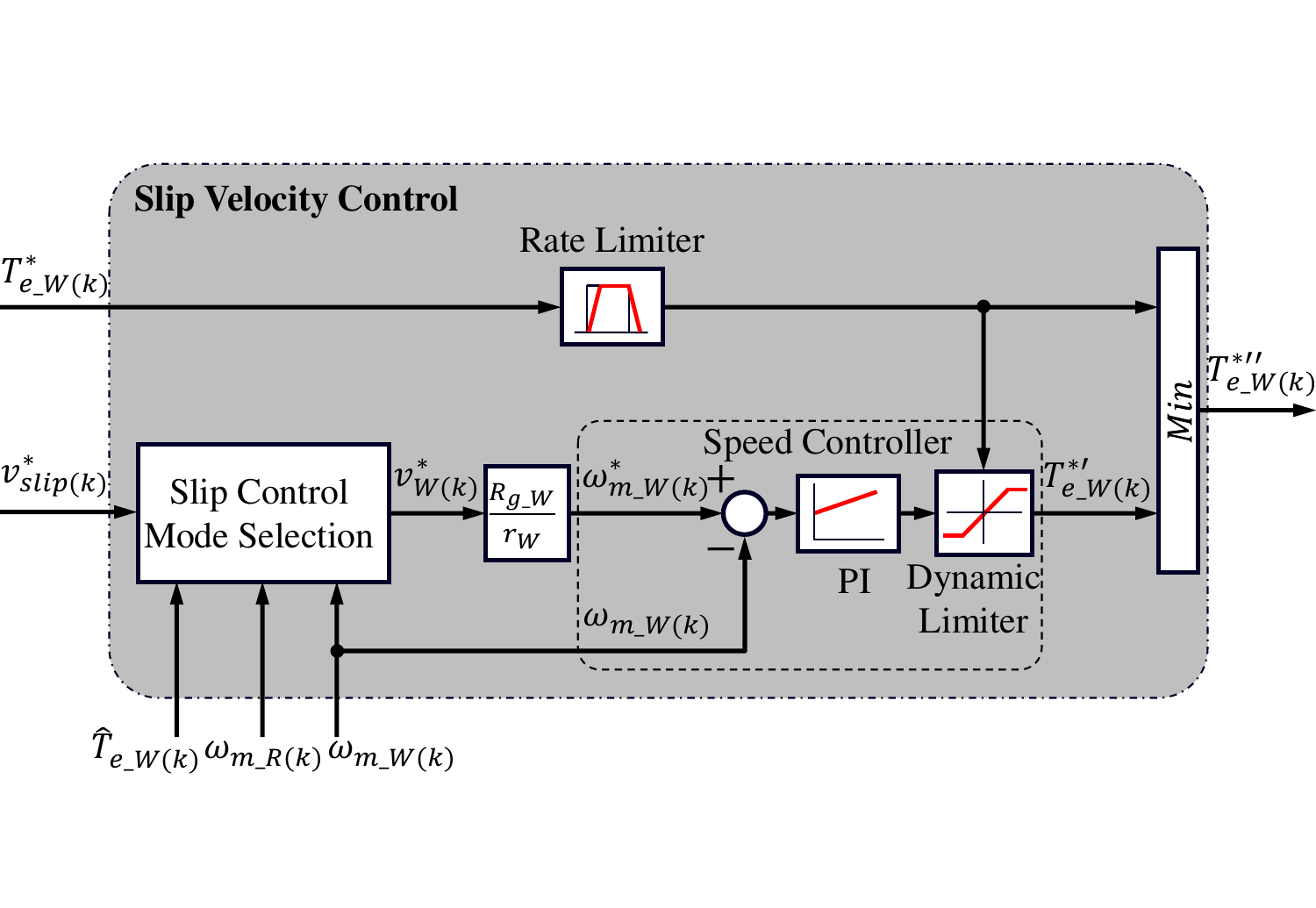}
    \caption{Detailed slip velocity control block diagram. Slip Control Mode Selection block is explained in Sections \ref{sec:overview_slip_control} and \ref{sec:proposed}.}
    \label{fig:slip_control_all}
\end{figure}
For the following discussion and simulations, it is assumed that the roller drive control is set to speed control mode running with a constant speed. This is consistent with the case of a train that has a very large inertia. Meanwhile, the wheel drive operates with torque control; the slip velocity controller remains disabled unless the actual slip surpasses the established limit (see Fig. \ref{fig:slip_control_all}). The torque reference $T^{\ast}_{e\_W}$ is transferred to the wheel drive torque command, i.e, $T^{\ast\prime\prime}_{e\_W}=T^{\ast}_{e\_W}$ unless wheel slip is detected. In this case, the slip velocity control is activated, and the torque command generated from the slip speed controller is passed by the ($Min$) function, i.e., $T^{\ast\prime\prime}_{e\_W}=T^{\ast\prime}_{e\_W}$. Wheel speed reference $\omega^{\ast}_{m\_W}$ is obtained from the slip velocity control block considering the gear ratio, i.e. $\omega^{\ast}_{m\_W}= \frac{R_{g\_W}}{r^{}_W} \cdot v^{}_W$, where $v^{}_W$ is the wheel linear speed in m/s. A dynamic limiter is added for the slip speed controller to avoid wind-up problems in the Propositional-Integral (PI) controller when the slip speed control is not active \cite{nihal2022design}. Different slip control modes are discussed following.

\section{Overview of Wheel-Rail Slip Velocity Control Methods} \label{sec:overview_slip_control}

This section reviews the slip velocity control methods reported in the literature. Simulation results using the down-scaled test rig are provided. Experimental results will be shown in section \ref{sec:Exp}. 

\begin{figure}[!h]
    \centering
    \includegraphics[width=0.9\columnwidth, trim={0cm 5cm 0cm 0cm},clip]{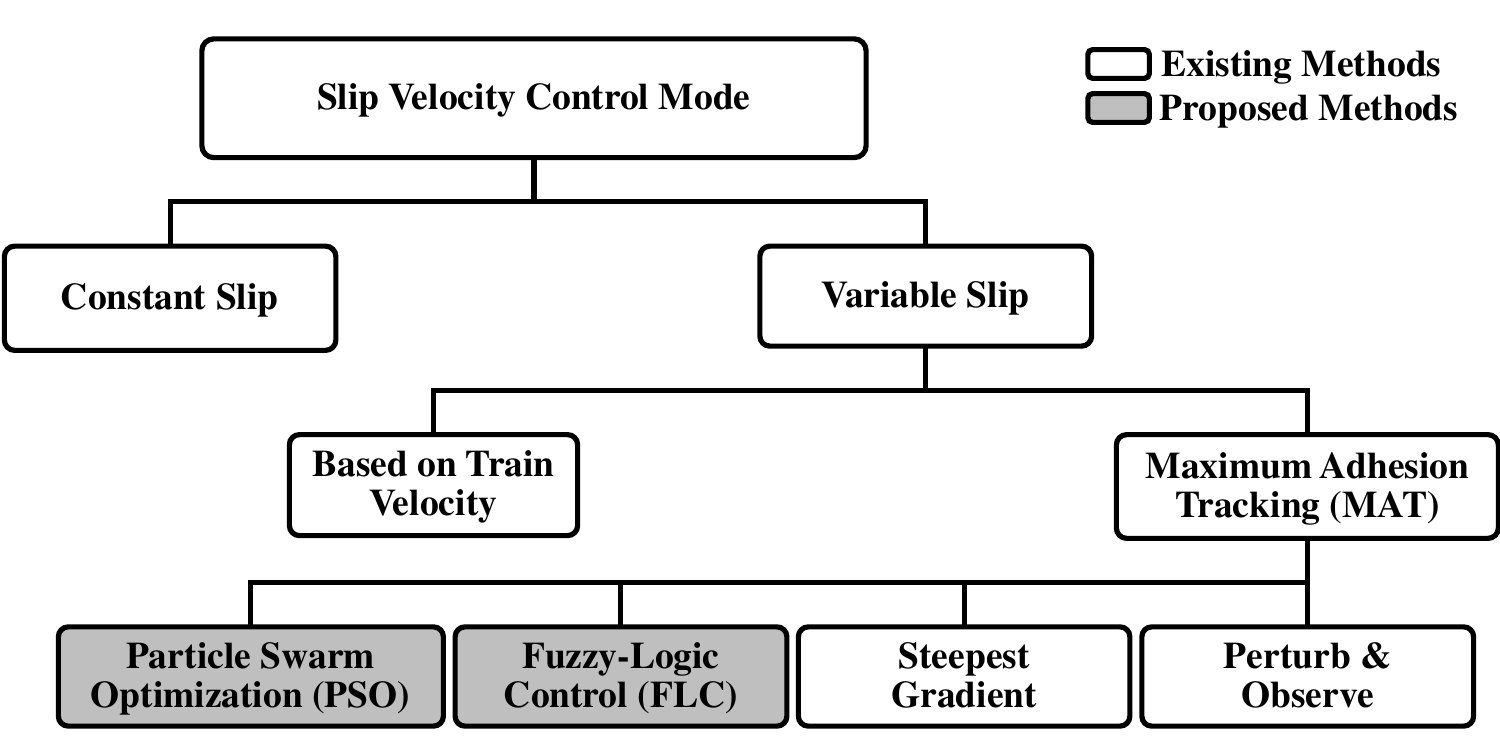}
    \caption{Classification of slip velocity control mode.}
    \label{fig:slip_mode}
\end{figure}

The slip velocity control mode in Fig. \ref{fig:slip_control_all} can be selected either with constant or variable slip velocity. Variable slip mode can be based on train speed where the slip velocity reference value is changing along the whole trip, and continuously adapting the slip velocity command to track the maximum adhesion. The classification of the slip control mode is summarized in Fig. \ref{fig:slip_mode}, and the discussion of each method is provided in the following subsection. 

\subsection{Constant slip velocity control}  \label{subsec:constant_slip_control}

\begin{figure}[t]
    \centering
    \includegraphics[width=0.9\columnwidth, trim={3cm 4cm 3cm 7.5cm},clip]{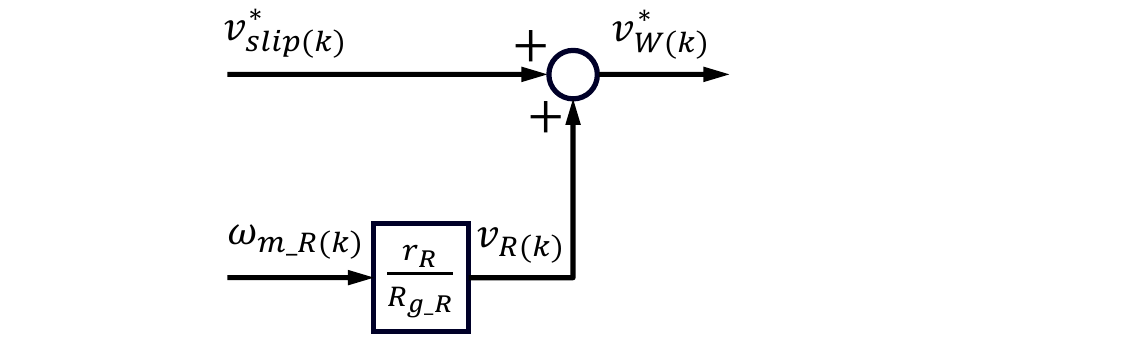}
    \caption{Constant slip velocity control command generation.}
    \label{fig:constant_slip}
\end{figure}
\begin{figure}[t]
    \centering
        \subfigure[]
    {
            \includegraphics[width=0.8\columnwidth,trim={0cm 0cm 0cm 0cm},clip]{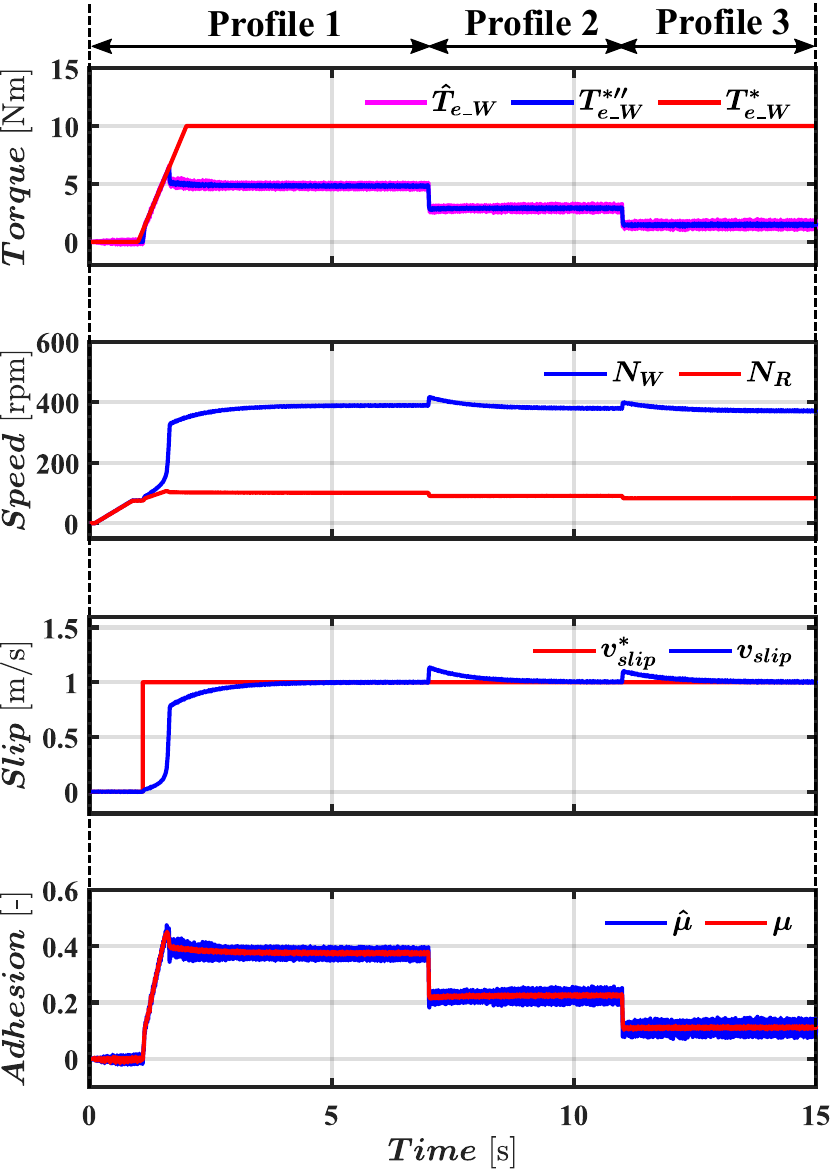}
        \label{fig:time_const}
    }
    \subfigure[]
    {
    \includegraphics[width=0.85\columnwidth, trim={0cm 0cm 0cm 0cm},clip]{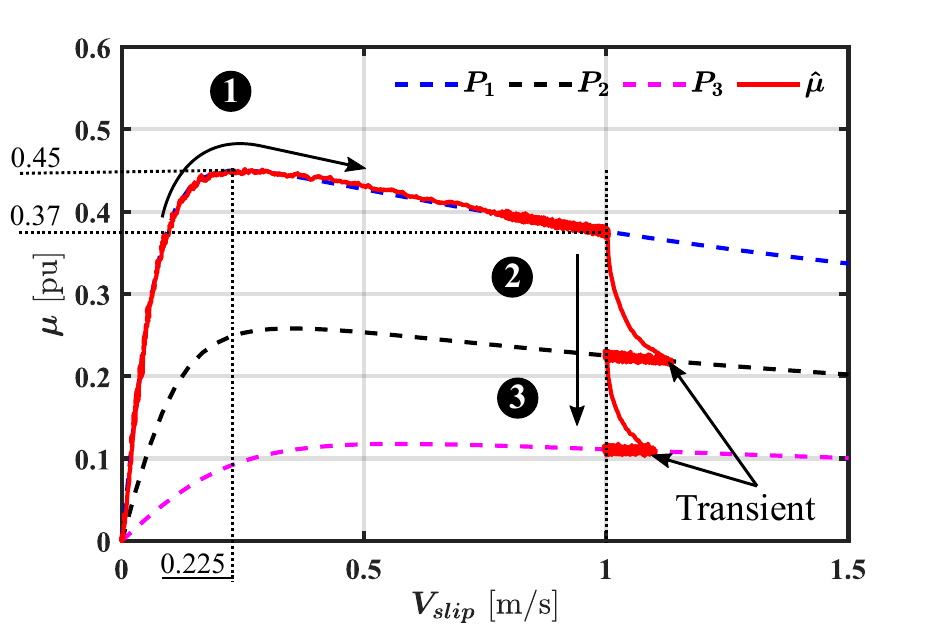}
    \label{fig:mu_const}
    } 
    \caption{Constant slip velocity control (simulation): (a) transient response. 1- From top to bottom: Wheel torque command, limited torque, estimated torque; 2-  wheel and roller speeds; 3- slip command and actual slip; 4- estimated and actual adhesion; (b) adhesion profiles. \textcircled{1} $P_1: t< 7$ s; \textcircled{2} $P_2: 7$~s$ < t < 11$ s; \textcircled{3} $P_3: t>11$~s.}
    \label{fig:result_const}
\end{figure}
This method is the simplest solution for slip control, being likely the most common choice for anti-slip protection in railway applications \cite{pichlik2014overview}. In this control mode, slip velocity reference $v^{\ast}_{slip}$ is added to the train velocity and sent to the wheel speed controller as seen in Fig. \ref{fig:constant_slip}. In this paper, it is assumed that the train speed is measured. Methods to measure or estimate the train's linear speed can be found in \cite{malvezzi2001train, mei2008novel, guzinski2009speed}. Slip velocity reference is selected based on field tests and the train's driver experience, being generally in the range of $\approx$ 0.5 to 1.0 m/s \cite{pichlik2014overview}. Regardless of its simplicity, the main demerit of this method is that it does not usually operate at the maximum adhesion point.

Fig. \ref{fig:time_const} shows simulation results using this method. The adhesion profiles used are shown in Fig. \ref{fig:mu_const}. A torque ramp with a final value of 10 Nm is commanded. The adhesion is lost when the actual torque reaches $\approx 5$ Nm (see Fig. \ref{fig:time_const}-top). Slip velocity control is then activated with a 1.0 m/s set point (see third subplot in Fig. \ref{fig:time_const}). The response of slip control to changes in the wheel-roller adhesion conditions is also simulated as shown in Fig. \ref{fig:result_const}. 
 Constant slip velocity control is seen to provide a good dynamic response, with deviations from the target sleep corrected in $\approx 1$ s, and not exceeding $\approx 0.2$ m/s. 

\subsection{Variable slip velocity control with maximum adhesion estimation}  \label{subsec:variable_slip_control}

These methods are aimed to operate at the slip speed providing maximum adhesion. This will require maximum adhesion estimation. Subsections \ref{subsec:PO_slip_control} and \ref{subsec:steepes_slip_control} discuss methods already reported in the literature. Two new methods will be proposed in Section \ref{sec:proposed}

\subsubsection{Perturb and Observe (P\&O)}  \label{subsec:PO_slip_control}

\begin{figure}[tb]
    \centering
    \includegraphics[width=\columnwidth, trim={0cm 2cm 0cm 0cm},clip]{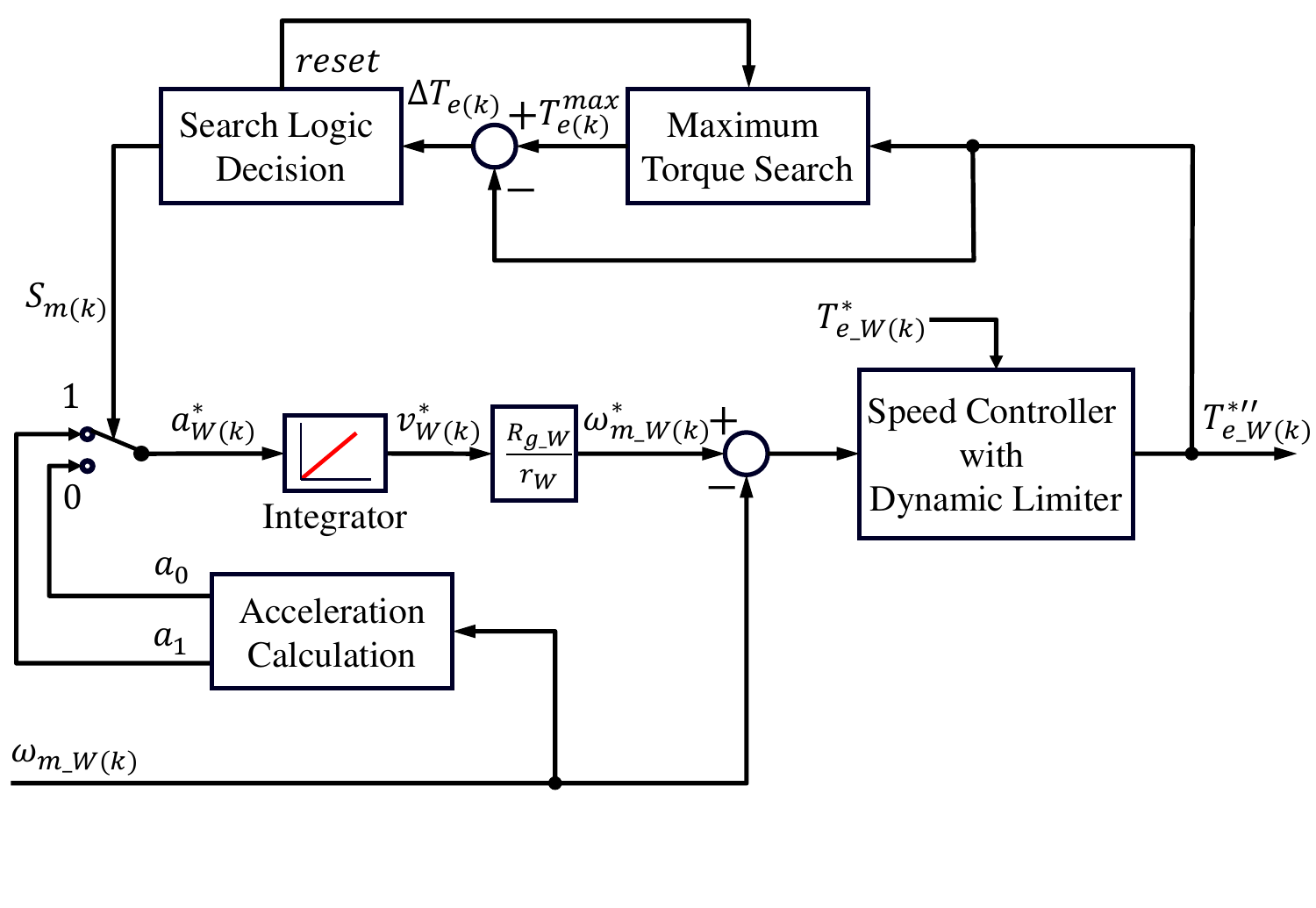}
    \caption{Perturb and Observe (P\&O) slip velocity control mode block diagram.}
    \label{fig:PO}
\end{figure}

In this method, the slip velocity is indirectly controlled by perturbing the wheel acceleration  as seen in Fig. \ref{fig:PO}. The wheel velocity command $v^{\ast}_W$ is obtained by integrating the wheel acceleration command which is a combination of the current acceleration $a^{}_W$ and a constant value $\Delta a$ as given by \eqref{eq:a0} and \eqref{eq:a1}. Selection of $a_0$ or $a_1$ is based on Perturb and Observe (P\&O) technique to track the maximum torque \cite{schwartz1992regelung,buscher1995radschlupfregelung}.
    
\begin{align} 
        a^{}_0 &= a^{}_W - \Delta a \label{eq:a0} \\
        a^{}_1 &= a^{}_W + \Delta a \label{eq:a1} 
\end{align}

The operation of MAT strategy can be summarized as follows: 
\begin{itemize}
    \item The acceleration of the traction motor wheel is initially perturbed (e.g. increased). The developed torque during this process is stored. The maximum torque value during the perturbation period  $T^{max}_{e(k)}$ is held and subtracted from the current torque value obtaining $\Delta T_e$ (see Maximum Torque Search block in Fig. \ref{fig:PO}).
    \item $\Delta T_e$ is sent to the Search Logic Decision block, which will choose between $a^{}_0$ and $a^{}_1$ using a binary output signal $S_{m(k)}$. The search logic task is to adapt the operating point either by increasing or decreasing the acceleration command based on the current load torque compared to the maximum stored value during perturbation. If $\Delta T_e > T_{threshold}$, then the current torque $T^{\ast \prime \prime}_{e\_W}$ is moving apart from the peak of the adhesion curve thus the search logic block output $S_m=0$ to decelerate the wheel and bring the operating point back to the peak of the curve (see Signal Adaptation block in Fig. \ref{fig:PO}). 
    \item A reset signal generated in the Search Logic in Fig. \ref{fig:PO} is used to reset the counter in the Maximum Torque Search block which handles the perturbation period. Consequently, the operating point is expected to be alternating around the peak of the adhesion curve with no need for additional speed measurement, i.e. train velocity.  
\end{itemize}

\begin{figure}[tb]
    \centering
        \subfigure[]
    {
            \includegraphics[width=0.8\columnwidth,trim={0cm 0cm 0cm 0cm},clip]{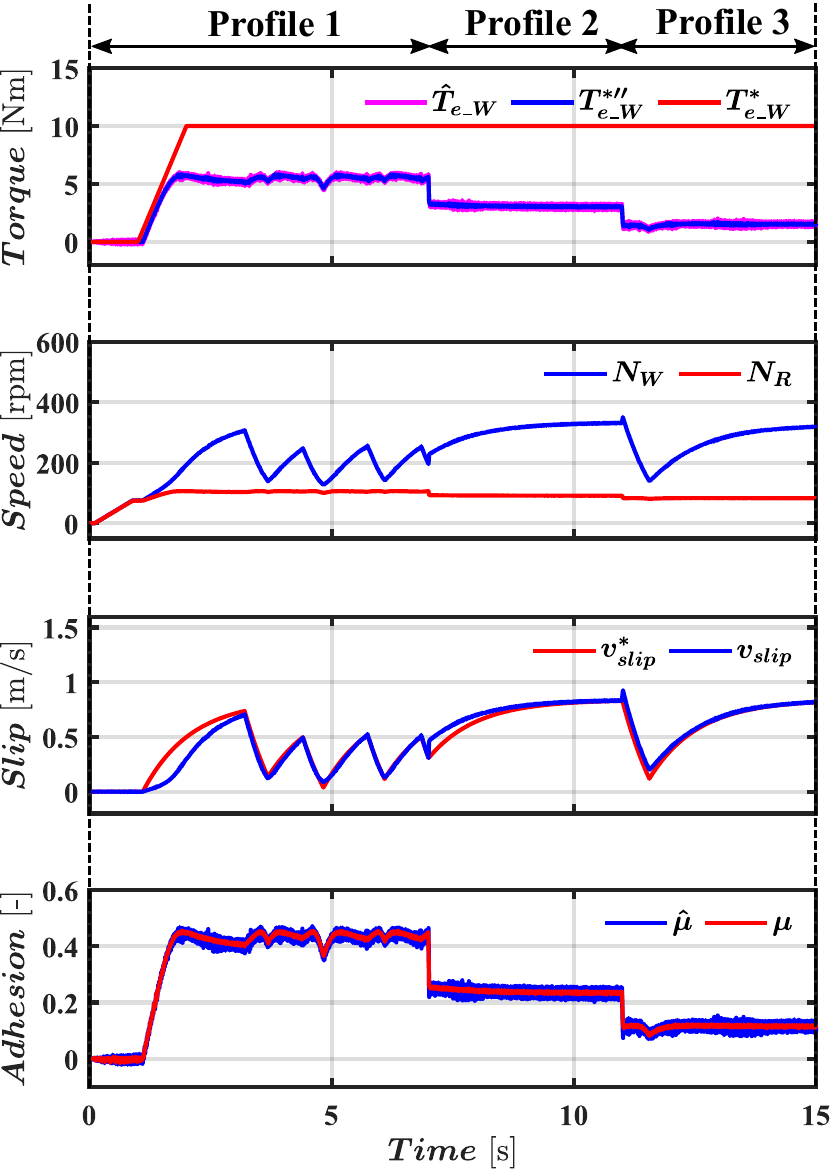}
        \label{fig:time_po}
    }
    \subfigure[]
    {
    \includegraphics[width=0.85\columnwidth, trim={0cm 0cm 0cm 0cm},clip]{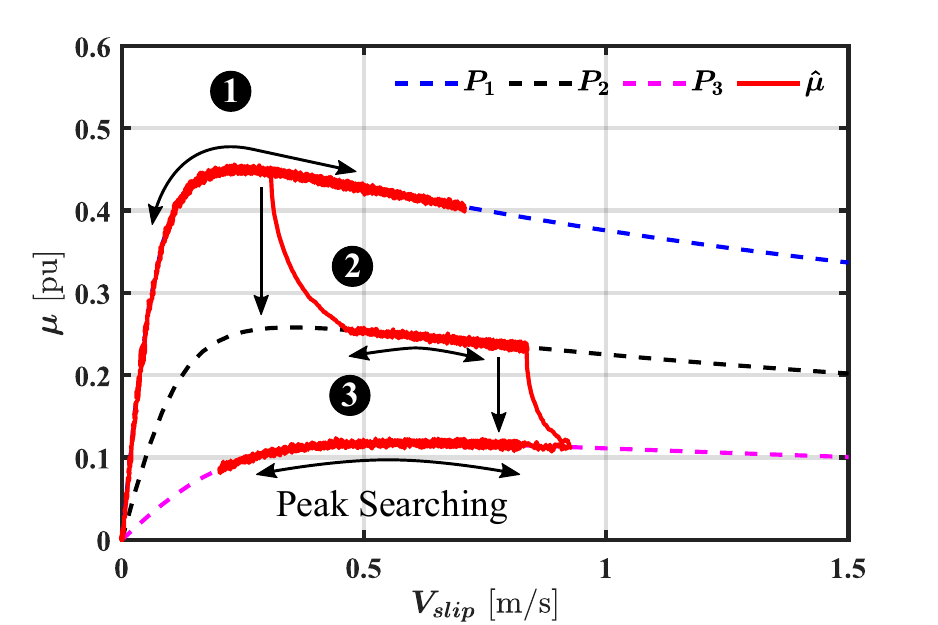}
    \label{fig:mu_po}
    } 
    \caption{Variable slip velocity control using Perturb and Observe (P\&O) (simulation): (a) transient response; (b) adhesion profiles. \textcircled{1} $P_1: t< 7$ s; \textcircled{2} $P_2: 7$ s$ < t < 11$ s; \textcircled{3} $P_3: t>11$~s.}
    \label{fig:result_po}
\end{figure}

The main drawback of this method is that it creates additional ripples in the machine torque, which depend on the perturbation period. This might contribute to undesired oscillations in the mechanical drive train torsional elements \cite{abouzeid2022torsional}.   
   
Fig. \ref{fig:result_po} shows the simulation results of the roller rig emulator using P\&O method slip velocity control. Operating conditions are the same as in Fig. \ref{fig:result_const}. Significant excursions around the peak of the adhesion curve are observed in Fig. \ref{fig:mu_po}, especially for the case of $P_1$ which corresponds to the highest adhesion level, i.e. dry condition. Also, it is found that the P\&O method has a slow dynamic response while searching for the peak. Finally, peak searching capability becomes more challenging with flat adhesion curves like $P_2$ and $P_3$ (see Fig. \ref{fig:mu_po}). 

\subsubsection{Steepest Gradient}  \label{subsec:steepes_slip_control}
\begin{figure}[t]
    \centering
        \subfigure[]
    {
            \includegraphics[width=0.8\columnwidth,trim={4.5cm 3.5cm 4cm 2.5cm},clip]{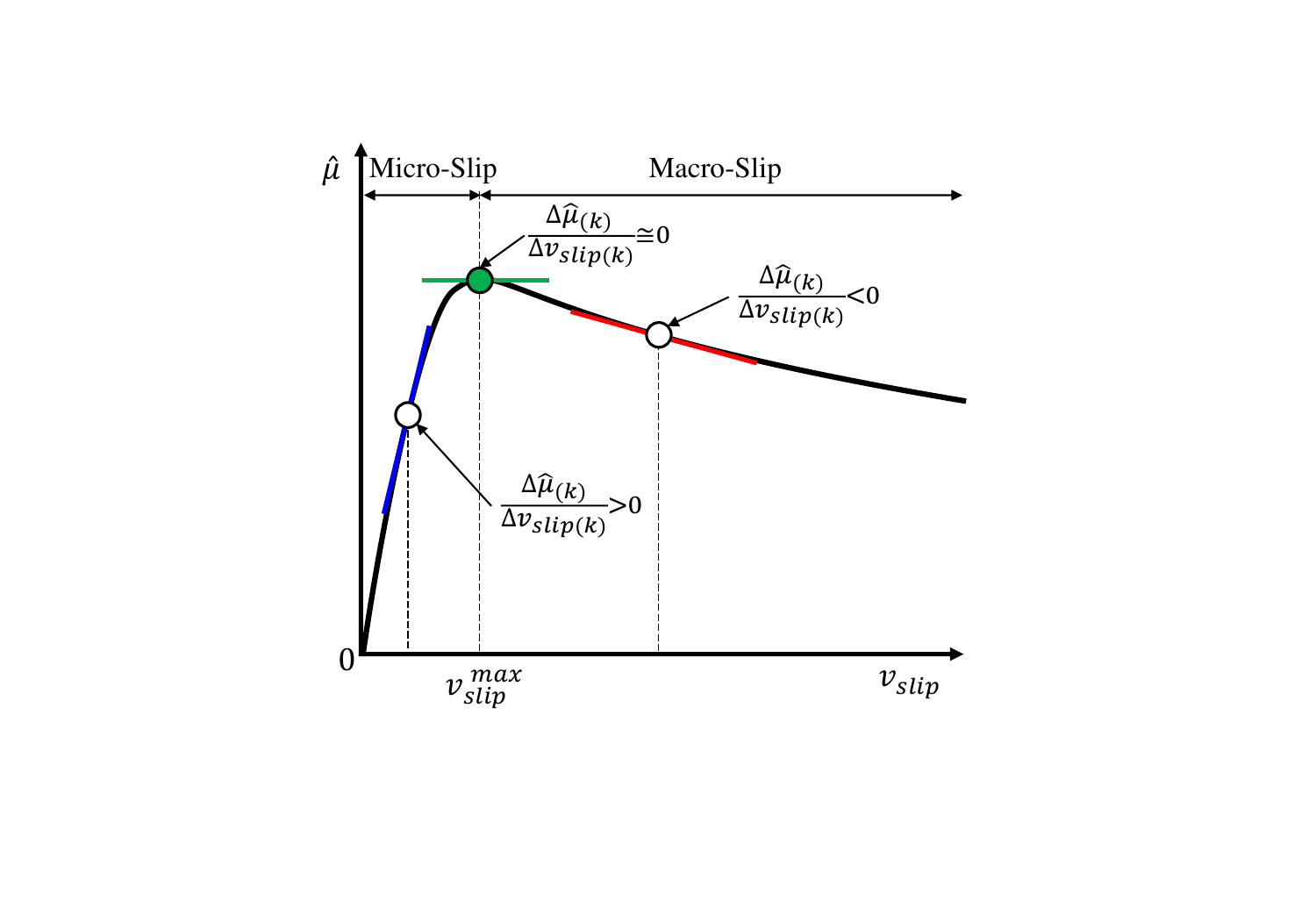}
        \label{fig:steepest_curve}
    }
    \subfigure[]
    {
    \includegraphics[width=\columnwidth, trim={0cm 3cm 0cm 0cm},clip]{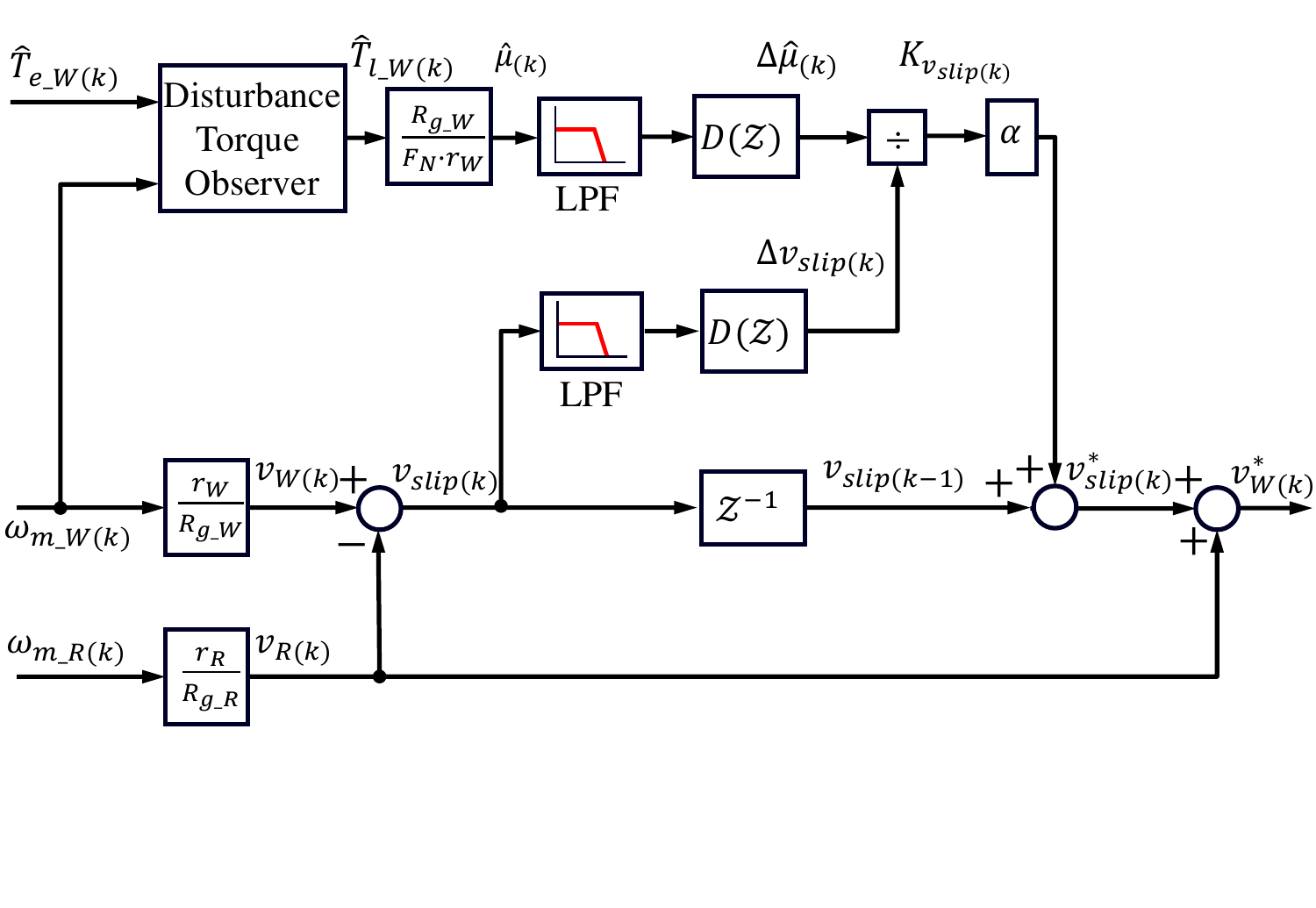}
    \label{fig:steepest_block}
    } 
    \caption{Steepest Gradient slip velocity control mode. (a) adhesion-slip curve and involved incremental variables; (b) Block diagram. }
    \label{fig:steepest}
\end{figure}

This approach takes advantage of the non-linear behavior of the adhesion-slip characteristic curve to track the maximum adhesion. As already known, the adhesion-slip characteristic is divided into two regions as shown in Fig. \ref{fig:steepest_curve}: 1) micro-slip (stable) region, where the adhesion coefficient $\mu$ increases with the slip velocity $v_{slip}$ till reaches its maximum value; 2) macro-slip (unstable), where any increase in slip velocity will decrease the adhesion coefficient and would drive the traction system to instability. 

Defining the increments of the adhesion coefficient and slip velocity as \eqref{eq:delta_mu_hat} and \eqref{eq:delta_vslip} respectively, the slope of the adhesion-slip curve is given by \eqref{eq:vslip_sg1}.
\begin{align}
v^{\ast}_{slip(k)} & = v^{}_{slip(k-1)}+\alpha \cdot K_{v_{slip}(k)}
\label{eq:vslip_sg} \\
\Delta \hat{\mu}_{(k)} & =\hat{\mu}_{(k)}-\hat{\mu}_{(k-1)} \label{eq:delta_mu_hat} \\
    \Delta {v_{slip(k)}} & =v_{slip(k)}-v_{slip(k-1)} \label{eq:delta_vslip}  \\
K_{v_{slip}(k)} & = \frac{\Delta \hat{\mu}_{(k)}}{\Delta v_{slip(k)}} 
\label{eq:vslip_sg1}
\end{align}

The slope $K_{v_{slip}(k)}$ is positive in the micro-slip region, and negative in the macro-slip region, being zero at the peak of the adhesion-slip curve [see Fig. \ref{fig:steepest_curve}].  
Hence, tracking the maximum adhesion in the steepest gradient method would be naturally done by adding the current gradient of the adhesion-slip curve to the previous slip velocity as shown in \eqref{eq:vslip_sg}. Gain $\alpha$ in \eqref{eq:vslip_sg} is an adaptation constant.

According to \eqref{eq:vslip_sg}, if the operating point is in the micro-slip region, the adhesion-slip gradient $K_{v_{slip}(k)}$ is positive and the slip velocity command is increased. Contrarily, if the operating point falls in the macro-slip region, the gradient added is negative, and the slip velocity command is decreased. Once the maximum adhesion is reached, the adhesion-slip gradient is zero and no change is applied to the slip velocity. The block diagram of the steepest gradient method is shown in Fig. \ref{fig:steepest_block}.
\begin{figure}[t]
    \centering
        \subfigure[]
    {
            \includegraphics[width=0.8\columnwidth,trim={0cm 0cm 0cm 0cm},clip]{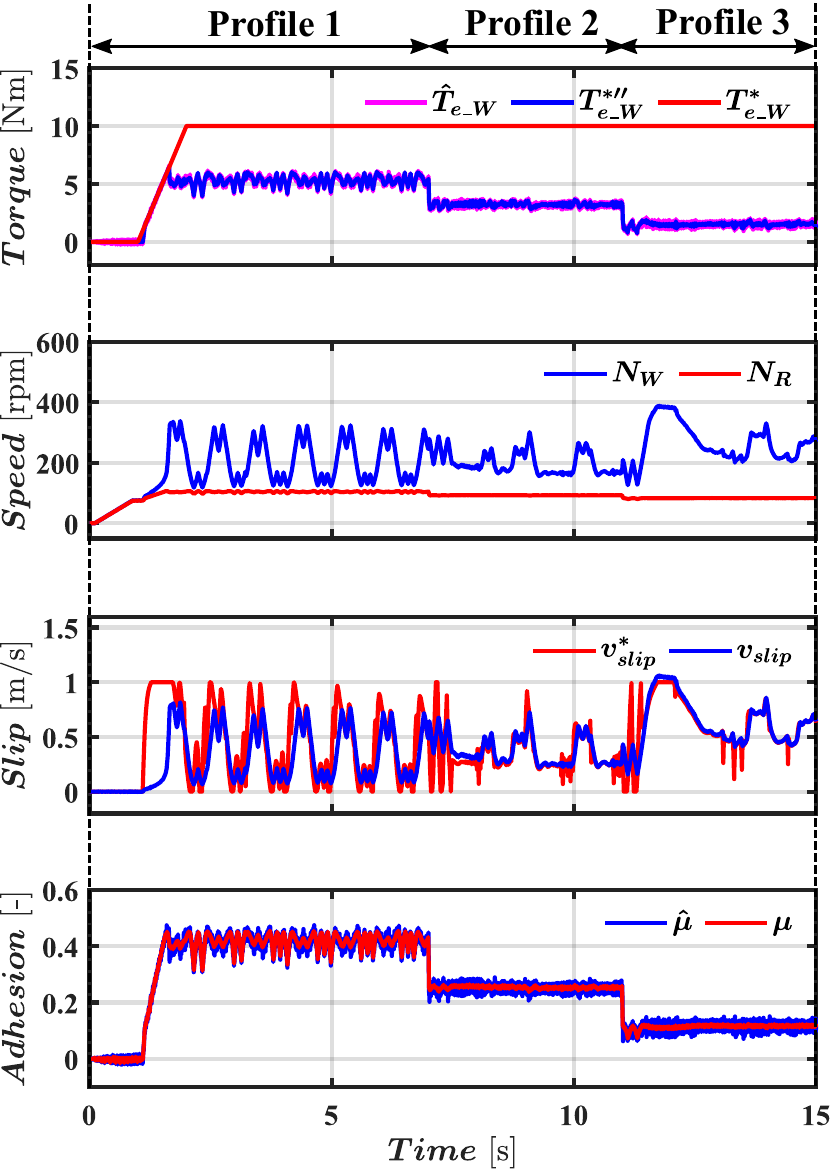}
        \label{fig:time_steepest}
    }
    \subfigure[]
    {
    \includegraphics[width=0.85\columnwidth, trim={0cm 0cm 0cm 0cm},clip]{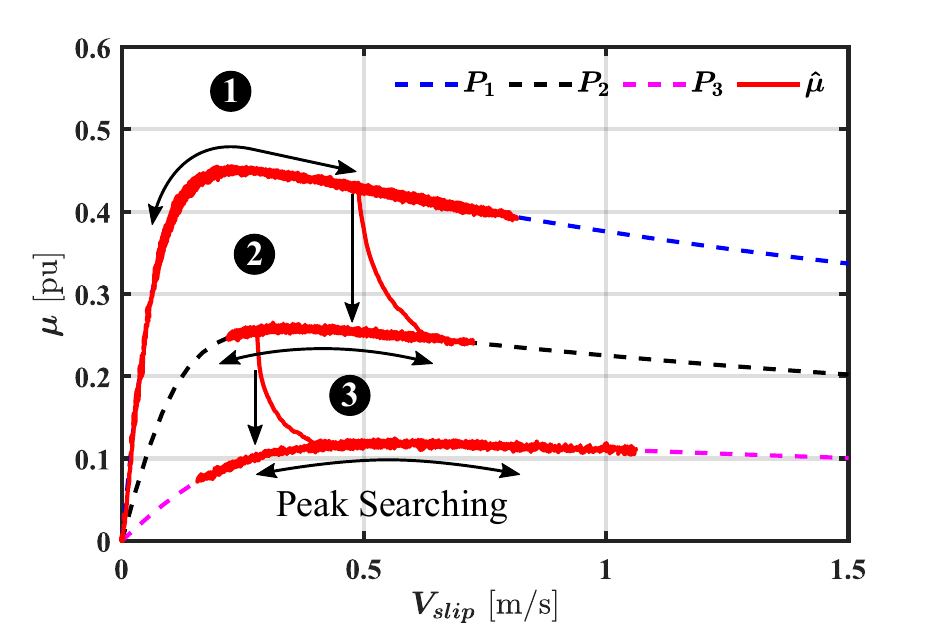}
    \label{fig:mu_steepest}
     } 
    \caption{Variable slip velocity control using Steepest Gradient method (simulation): (a) transient response; (b) adhesion profiles. \textcircled{1} $P_1: t< 7$ s; \textcircled{2} $P_2: 7$ s$ < t < 11$ s; \textcircled{3} $P_3: t>11$~s.}
    \label{fig:result_steepest}
\end{figure}
As the adhesion coefficient $\mu$ in an actual locomotive cannot be measured, estimation is required. A disturbance observer is commonly used for adhesion estimation using \eqref{eq:mu_hat}.

\begin{align} 
    \hat{\mu} &= \hat{T}_{l\_W}\frac{R_{g\_W}}{F_N\cdot r_W} \label{eq:mu_hat}
\end{align}

$\hat{T}_{l\_W}$ can be estimated from \eqref{eq:TL_diff}, where $\beta_m$ and $J_m$ are the viscus friction and inertia of the traction wheel motor, respectively.

\begin{align} 
\hat{T}_{l\_W} &=\hat{T}_{e\_W} -\beta_m \omega_{m\_W} - J_m \dot{\omega}_{m\_W}   \label{eq:TL_diff}
\end{align}

 To avoid the pure derivative in \eqref{eq:TL_diff}, a low pass filter can be used instead \eqref{eq:TlW}. 

\begin{align} 
\hat{T}_{l\_W(s)} &=\hat{T}_{e\_W(s)} - \omega_{m\_W(s)}\left[\beta_{m\_W} + J_m s \left(\frac{1}{\tau s + 1}\right)\right] \label{eq:TlW}
\end{align}

Two low-pass filters are used to attenuate the measurement noise in the slip velocity signal and the estimated adhesion coefficient prior to the discrete differentiation realized by $D(z)$ blocks. 

Simulation results for this method are shown in Fig. \ref{fig:result_steepest}. Due to the differentiation of signals used to estimate the adhesion slope $\frac{\Delta \hat{\mu_{(k)}}}{\Delta v_{slip(k)}}$, Steepest Gradient (SG) method suffers from high oscillations attempting to keep the operating point at the peak of the adhesion-slip curve [see Fig. \ref{fig:time_steepest}]. This becomes obvious for the adhesion profiles with higher slopes like $P_1$ as the resulting slope correction signal $K_{v_{slip}(k)}$ increases dramatically for the next step of the slip velocity command. Contrarily, with lower adhesion slopes such as $P_2$ and $P_3$, the correction signal moderately increases with the assumption of using constant correction factor $\alpha$. However, it can be noticed in this method that the operating points are closer to the peak of the adhesion-slip curves for all the profiles. However, the peak searching space is still high [see Fig. \ref{fig:mu_steepest}].

\section{Proposed MAT Techniques for Railways}  \label{sec:proposed}

As discussed in subsection \ref{subsec:variable_slip_control}, operation with variable slip will require estimation of maximum adhesion. Two new methods are proposed and assessed in this section, based on knowledge-based Fuzzy Logic Control and Particle Swarm Optimization respectively.   

\subsection{Proposed MAT Using Fuzzy Logic Control}  \label{subsec:MATFLC}

Fuzzy Logic Control (FLC) is a knowledge-based control technique that uses linguistic rules designed for complex, uncertain, and non-linear systems without requiring mathematical models and/or parameter estimation \cite{lee1990fuzzy, su1994adaptive}. FLC was first introduced for anti-lock braking systems (ABS) in railway traction applications to prevent wheel skid on the rail, resulting in high braking performance and consequently lower braking distance compared to conventional PID controller \cite{cheok1998fuzzy, cheok2000combined, lin2003self, mirzaei2005design}. Later, FLC concept has been extended for wheel slip prevention and speed profile tracking in electric trains \cite{jordan1995locomotive, garcia1997antislipping, khatun2001comparison, moaveni2022fuzzy}. The use of FLC for MAT is developed following.

\begin{figure}[b]
    \centering
    \includegraphics[width=\columnwidth, trim={0cm 3cm 0cm 3cm},clip]{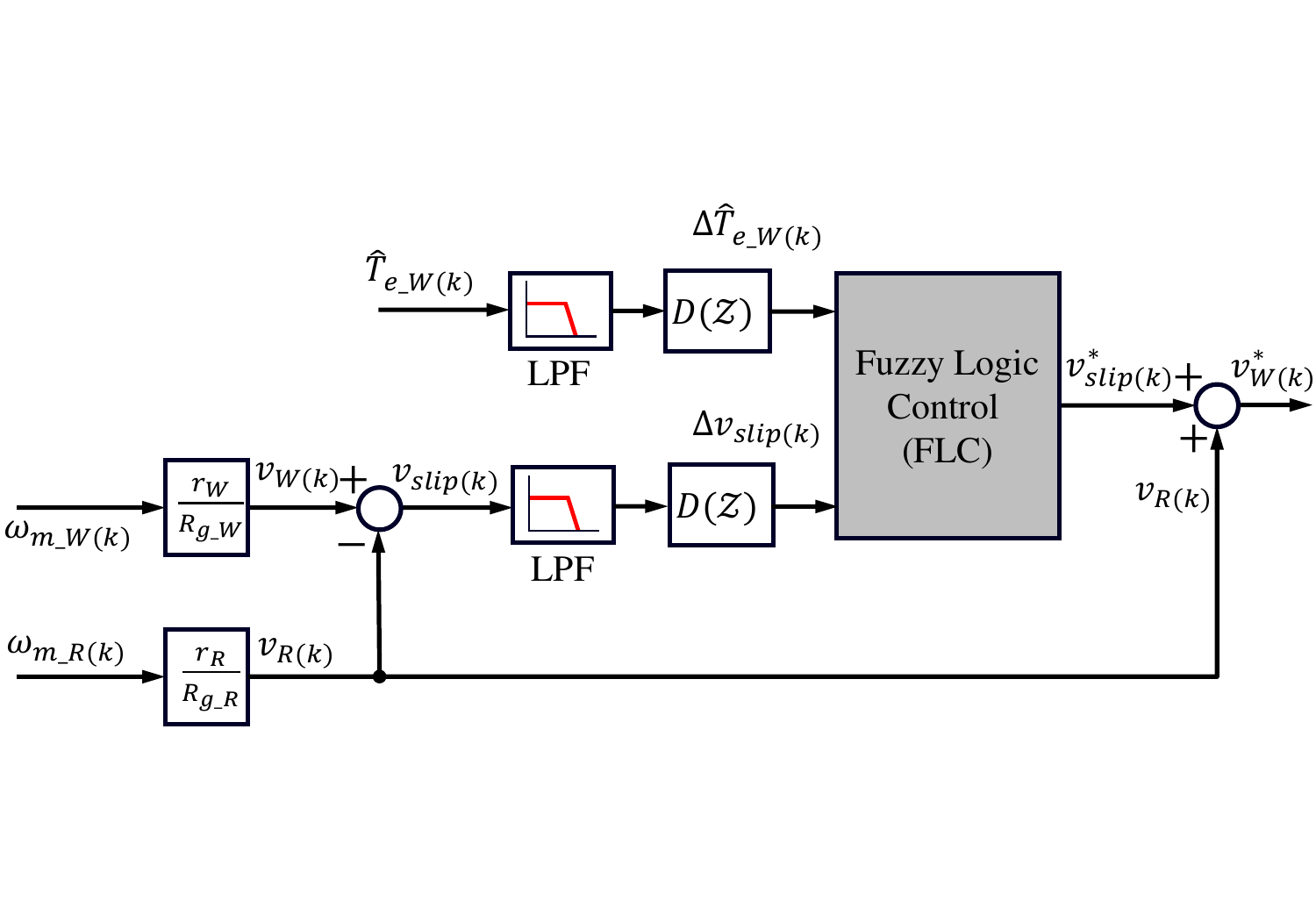}
    \caption{Proposed MAT using Fuzzy Logic Control (FLC) block diagram.}
    \label{fig:flc_block}
\end{figure}

\begin{figure}[ht]
    \centering
    \subfigure[]
    {
        \includegraphics[width=0.7\columnwidth,trim={4cm 2cm 4cm 3cm},clip]{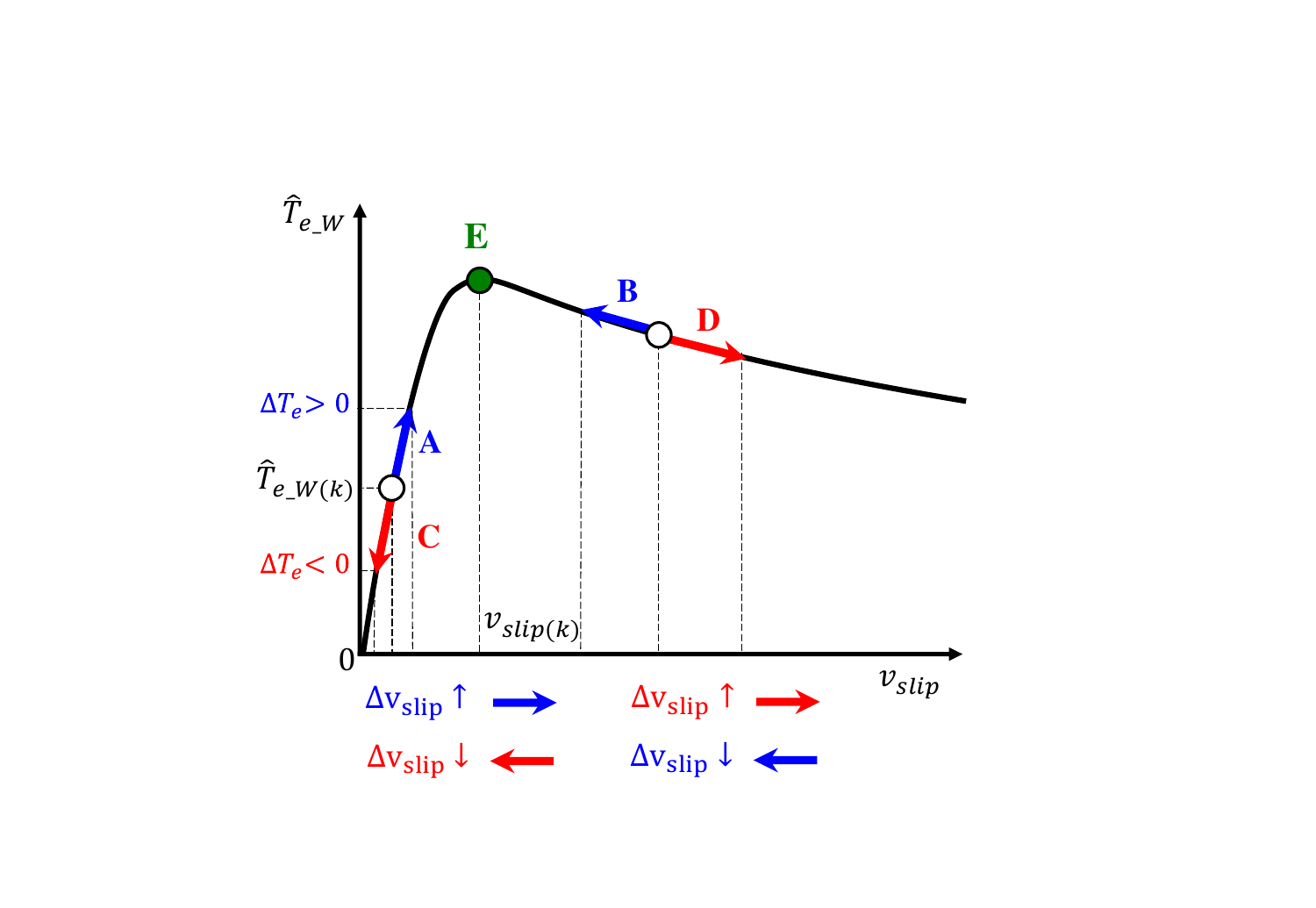}
        \label{fig:flc_curve}
    }
    \subfigure[]
    {
        \includegraphics[width=0.9\columnwidth,trim={0 0cm 0 0cm},clip]{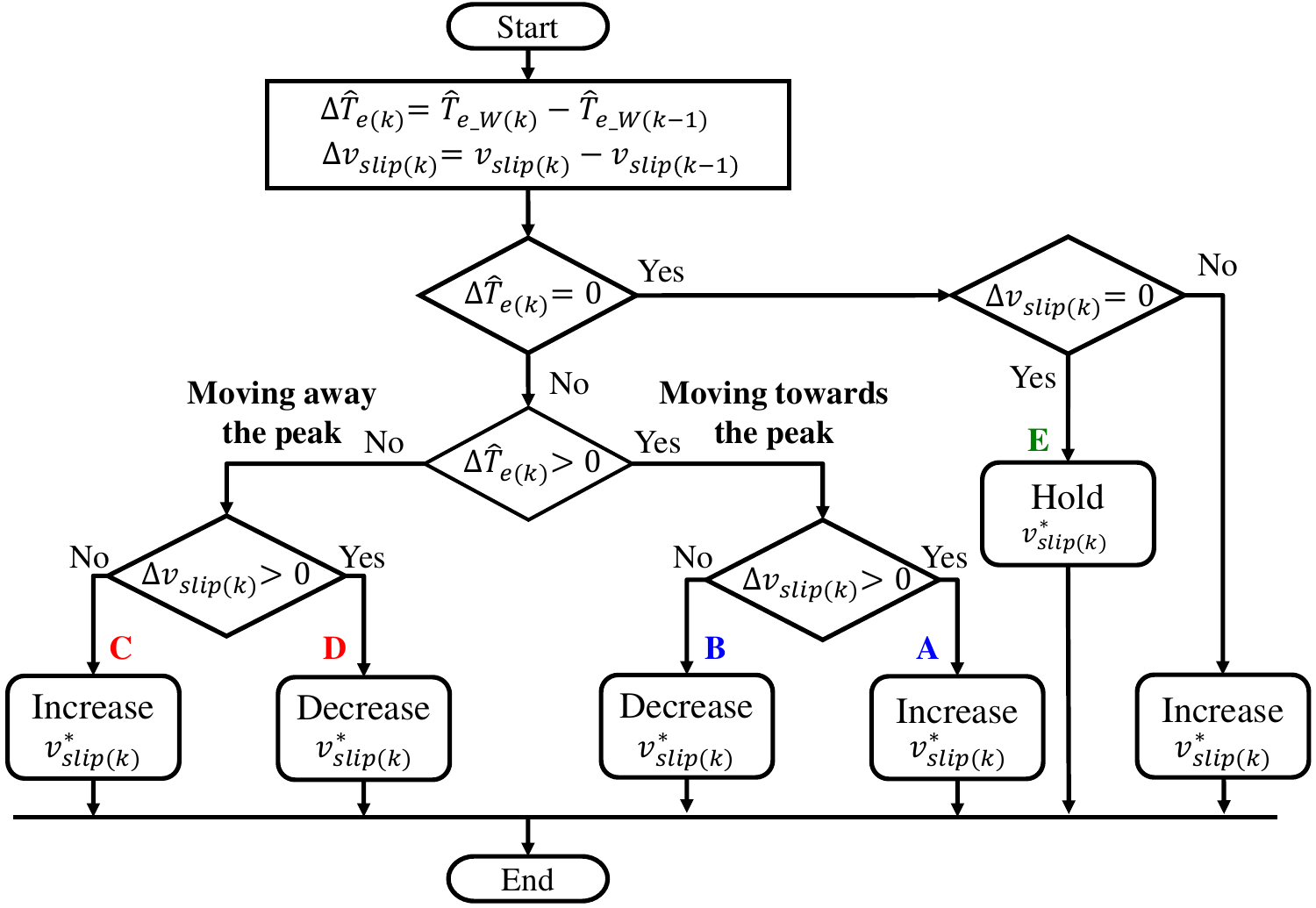}
        \label{fig:flc_flowchart}
    }
        \caption{Proposed MAT using FLC procedure. (a) Adhesion-slip curve strategy; (b) Flowchart representation.}
    \label{fig:flc_tracking}
\end{figure}
\begin{figure}[!h]
    \centering
    \subfigure[]
    {
        \includegraphics[width=0.9\columnwidth,trim={0cm 4cm 0cm 4cm},clip]{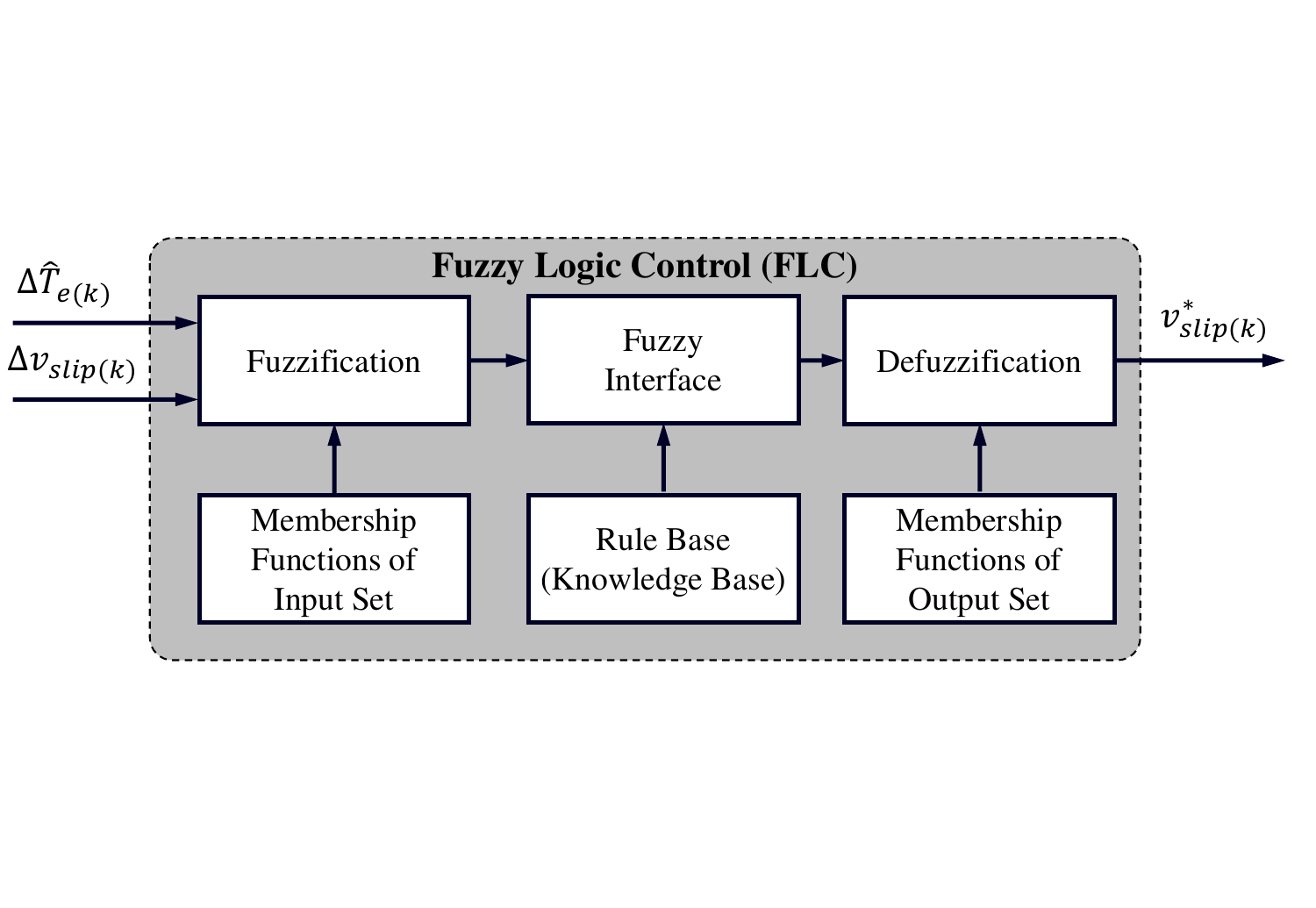}
        \label{fig:flc_detailed_blocks}
    }
    \subfigure[]
    {
        \includegraphics[width=0.9\columnwidth,trim={0 1cm 0 4cm},clip]{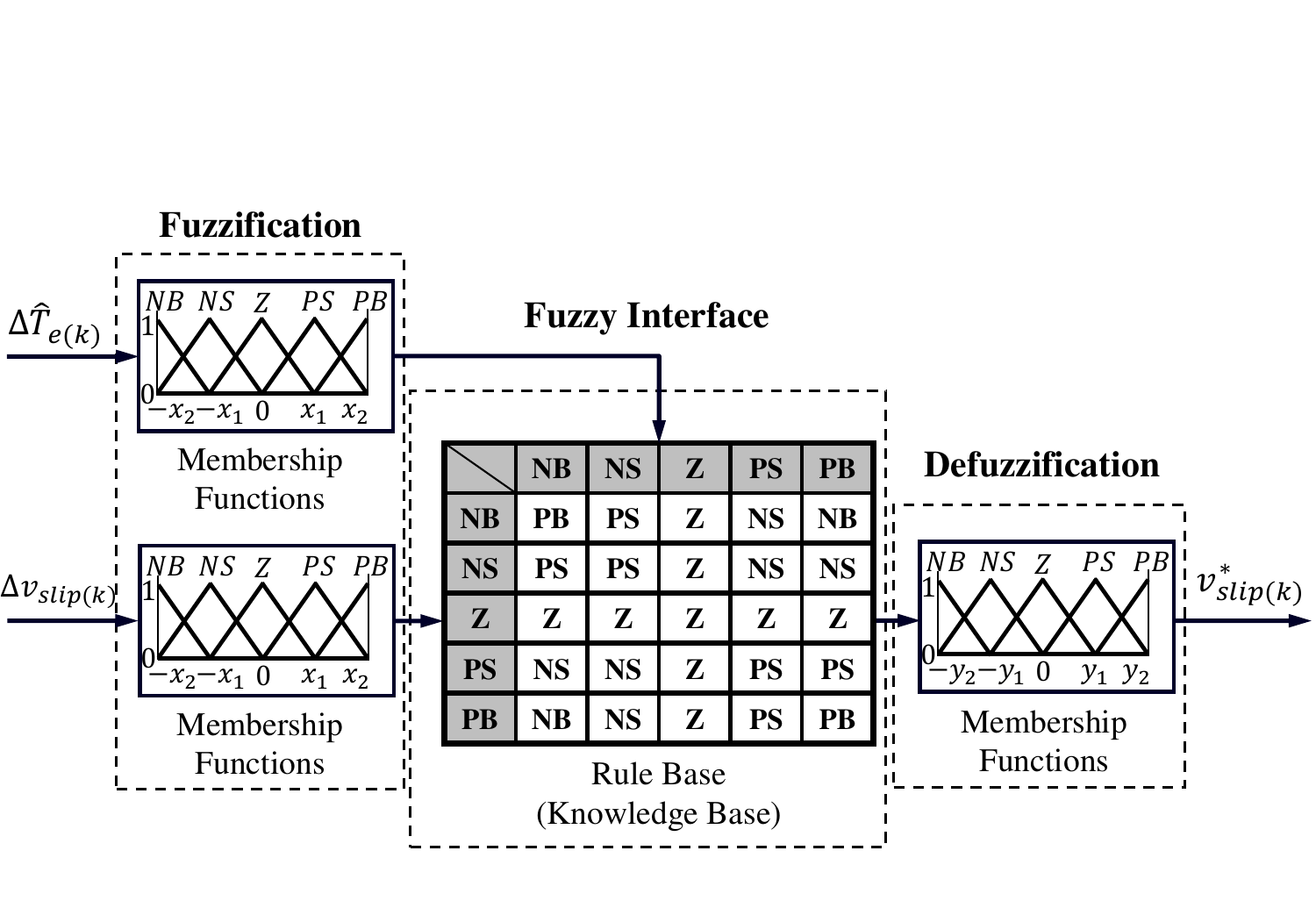}
        \label{fig:flc_membership}
    }
        \caption{Fuzzy Logic Control (FLC) scheme. (a) Basic FLC structure; (b) Input/Output Membership functions and Rules base for the proposed MAT-FLC.}
    \label{fig:flc_detailed}
\end{figure}

The proposed block diagram is shown in Fig. \ref{fig:flc_block}. It uses the same slip command adaptation concept of the Steepest Gradient method for tracking the peak of the adhesion curve [see Fig. \ref{fig:steepest_curve}]. However, the change in the slip command $v^{\ast}_{slip(k)}$ is adapted and generated automatically by the FLC block that uses the available torque and slip velocity signals which are available in the drive control scheme. From \eqref{eq:Tew}, it can be seen that the estimated wheel motor torque is proportional to the adhesion coefficient $\mu$ as the normal force $F_N$, the radius of the wheel $r_W$ and the gear ratio $R_{g_W}$ are already known. Thus the load torque estimation $\hat{T}_{l\_W}$ using the disturbance observer in Fig. \ref{fig:steepest_block} is not required anymore and the FLC rules can be applied to the estimated motor torque $\hat{T}_{e\_W}$ which is used for the drive torque control as shown in Fig.~\ref{fig:general_scheme}. 

The procedure for the proposed MAT using FLC is summarized in Fig.~\ref{fig:flc_tracking} including the flowchart shown in Fig. \ref{fig:flc_flowchart}.

\begin{figure}[t]
    \centering
        \subfigure[]
    {
            \includegraphics[width=0.8\columnwidth,trim={0cm 0cm 0cm 0cm},clip]{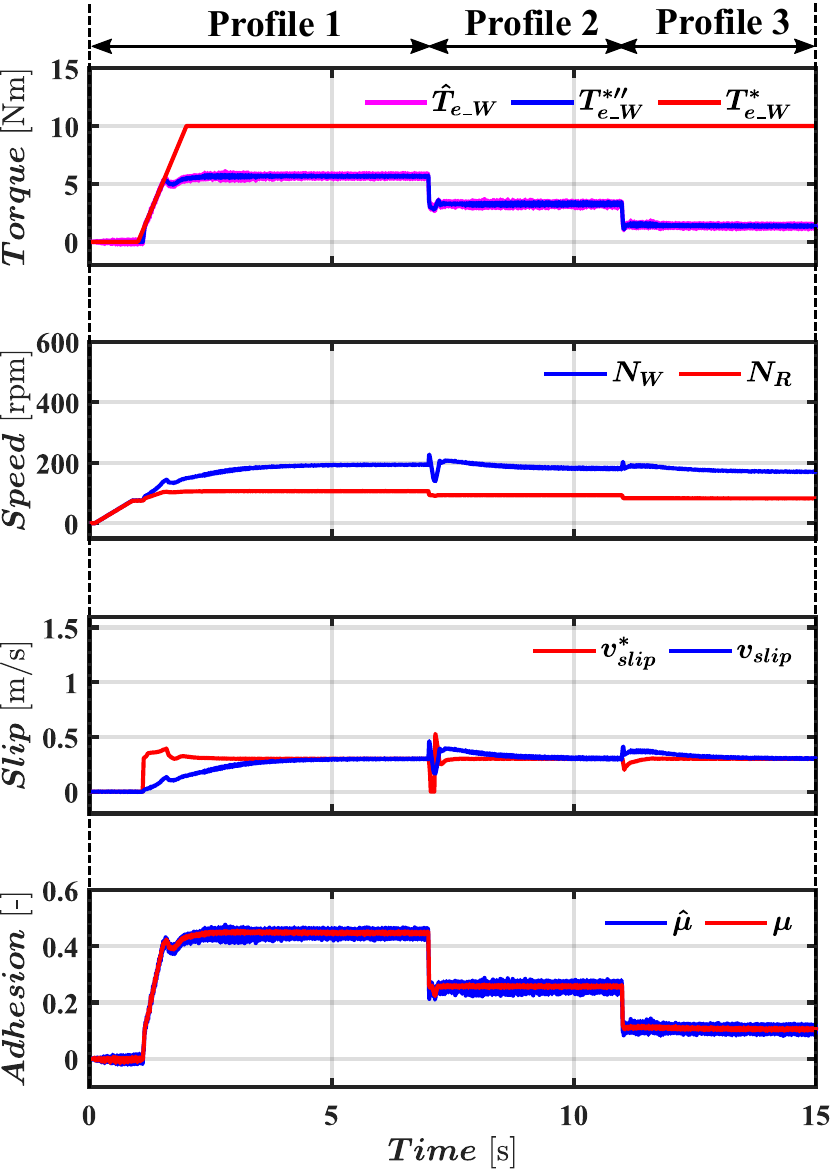}
        \label{fig:time_flc}
    }
    \subfigure[]
    {
    \includegraphics[width=0.85\columnwidth, trim={0cm 0cm 0cm 0cm},clip]{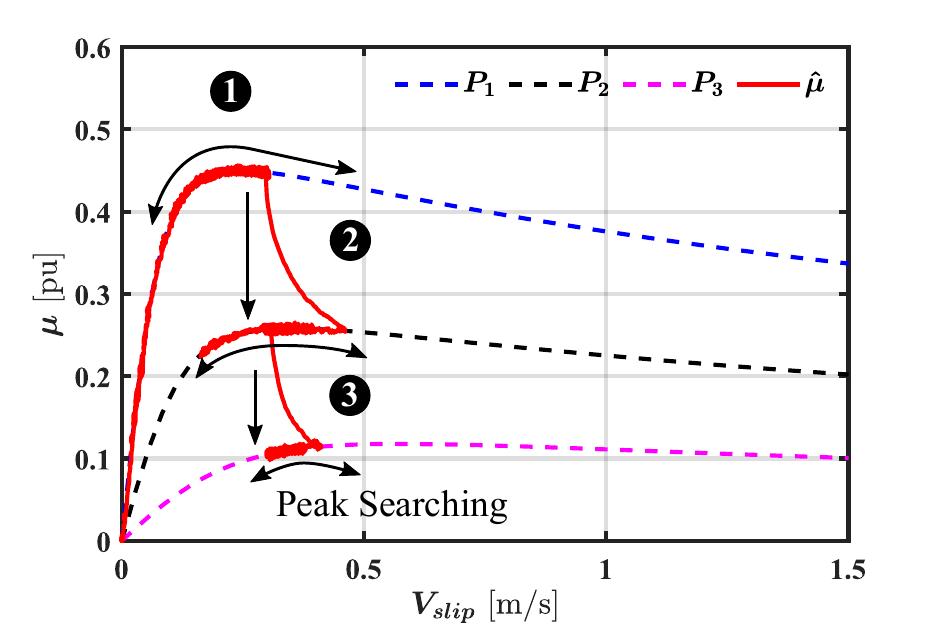}
    \label{fig:mu_flc}
    } 
    \caption{MAT using the proposed MAT-FLC (simulation): (a) transient response; (b) adhesion profiles. \textcircled{1} $P_1: t< 7$ s; \textcircled{2} $P_2: 7$ s$ < t < 11$ s; \textcircled{3} $P_3: t>11$~s.}
    \label{fig:result_flc}
\end{figure}
Like the conventional FLC structure, the FLC block of the proposed method (see Fig. \ref{fig:flc_block}) contains the input fuzzification, the fuzzy interface, and the output defuzzification respectively as seen in \ref{fig:flc_detailed_blocks}. The fuzzification block consists of two membership functions that create the linguistic rules of the input variables, i.e. the $\Delta \hat{T}_{e(k)}$ and $\Delta v_{slip(k)}$. Then, the fuzzy interface correlates the linguistic rules with the knowledge base for maximum adhesion-slip curve tracking. Finally, the fuzzified output rules are transformed back to real numbers using the defuzzification membership function. The rules used in \ref{fig:flc_membership} are denoted for: NB is Negative Big; NS is Negative small; Z is zero; PS is Positive Small, and PB is Positive Big. The choice of the input and output parameters ($x_2, x_1 \:\&\: y_2, y_1$) will depend on field tests and trains’ driver expertise. In this paper, the input parameters are assigned as $x_2=15\:\text{Nm},\: x_1=x_2/2=7.5\:\text{Nm}$ for torque increment, $x_2=1.0\:\text{m/s},\:x_1=x_2/2=0.5\:\text{m/s}$ for slip velocity increment, and the output parameters are $y_2=0.5\:\text{m/s},\:y_1=y_2/2=0.25\:\text{m/s}$ for the updated slip velocity command.     

Simulation results of the proposed MAT-FLC method are shown in Fig. \ref{fig:result_flc}. The improved dynamic response and reduced oscillations are readily visible comparing Fig. \ref{fig:time_flc} with Fig. \ref{fig:time_po} and \ref{fig:time_steepest}. The maximum adhesion of $\hat{\mu}=0.45$ for $P_1$ is achieved in $<1$s. The searching space is also decreased as observed comparing Fig. \ref{fig:mu_flc} with Fig. \ref{fig:mu_po} and \ref{fig:mu_steepest}.

For adhesion profiles $P_1$ and $P_2$, MAT-FLC was able to find the maximum adhesion-slip point. However, for $P_3$, the maximum adhesion found by the algorithm was $\approx 12\%$ smaller than the optimal value, with an error of $\approx -40\%$ in the estimated optimal slip velocity. This error can be minimized by modifying the membership functions and rules used in the FLC. Thus, adaptive tuning of FLC for multiple adhesion profiles to track the peak of the adhesion curve is needed. Implementing adaptive tuning algorithms increases the complexity of the proposed MAT-FLC \cite{shi1999empirical, herrera1995tuning, shi2000new,guenounou2014adaptive,patcharaprakiti2005maximum}. A new approach for MAT estimation that overcomes this problem is proposed in the next subsection.

\subsection{Proposed MAT Using Particle Swarm Optimization}  \label{subsec:MATPSO}
Particle swarm optimization (PSO) is a population-based stochastic optimization algorithm inspired by the movement of organisms such as flocks of birds or schools of fish \cite{kennedy1995particle}. PSO concept has roots in artificial life and evolutionary computation, intended for optimizing non-linear functions \cite{poli2007particle, wang2018particle}. PSO algorithm is simple, computationally efficient, and effective in solving a variety of problems for different applications \cite{shi1999empirical, shi2001particle}. Maximum Power Point Tracking using Particle swarm optimization (MPPT-PSO) is considered one of the most popular evolutionary optimization algorithms in solar Photo-Voltaic (PV) systems due to its high tracking speed, ability to operate under different environmental conditions, and fast computational capability \cite{liu2012particle, ishaque2012improved, li2018overall, xu2020improved, diaz2021evaluation, verma2021meta}.

\begin{figure}[b]
    \centering
    \includegraphics[width=\columnwidth, trim={0cm 3cm 0cm 3cm},clip]{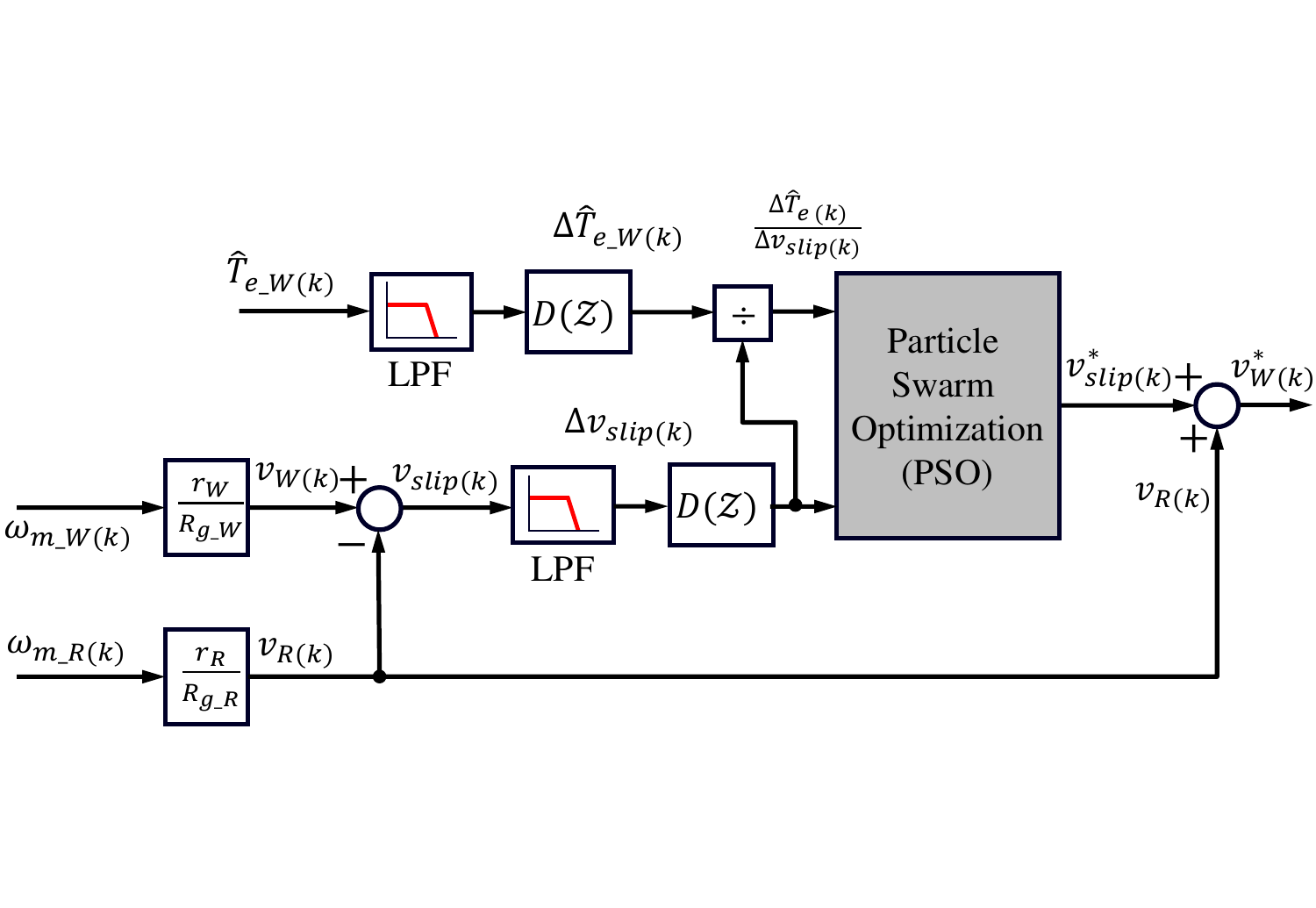}
    \caption{Proposed MAT using Particle Swarm Optimization (PSO) block diagram.}
    \label{fig:pso_slip}
\end{figure}

\begin{figure}[!h]
    \centering
    \subfigure[]
    {
        \includegraphics[width=0.4\textwidth,trim={4cm 4cm 4cm 4cm},clip]{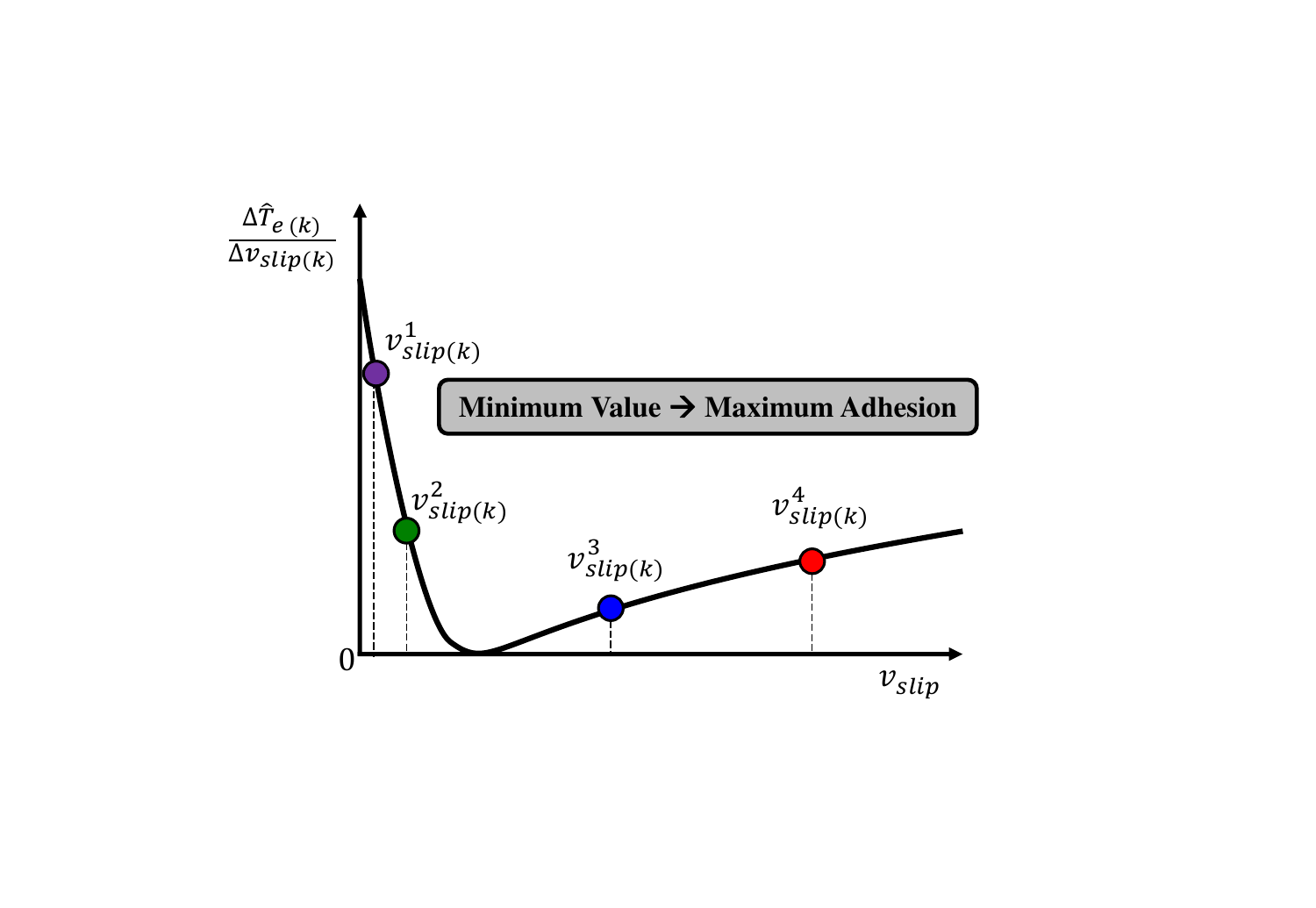}
        \label{fig:pso_0}
    }
    \subfigure[]
    {
        \includegraphics[width=0.4\textwidth,trim={4cm 4cm 4cm 4cm},clip]{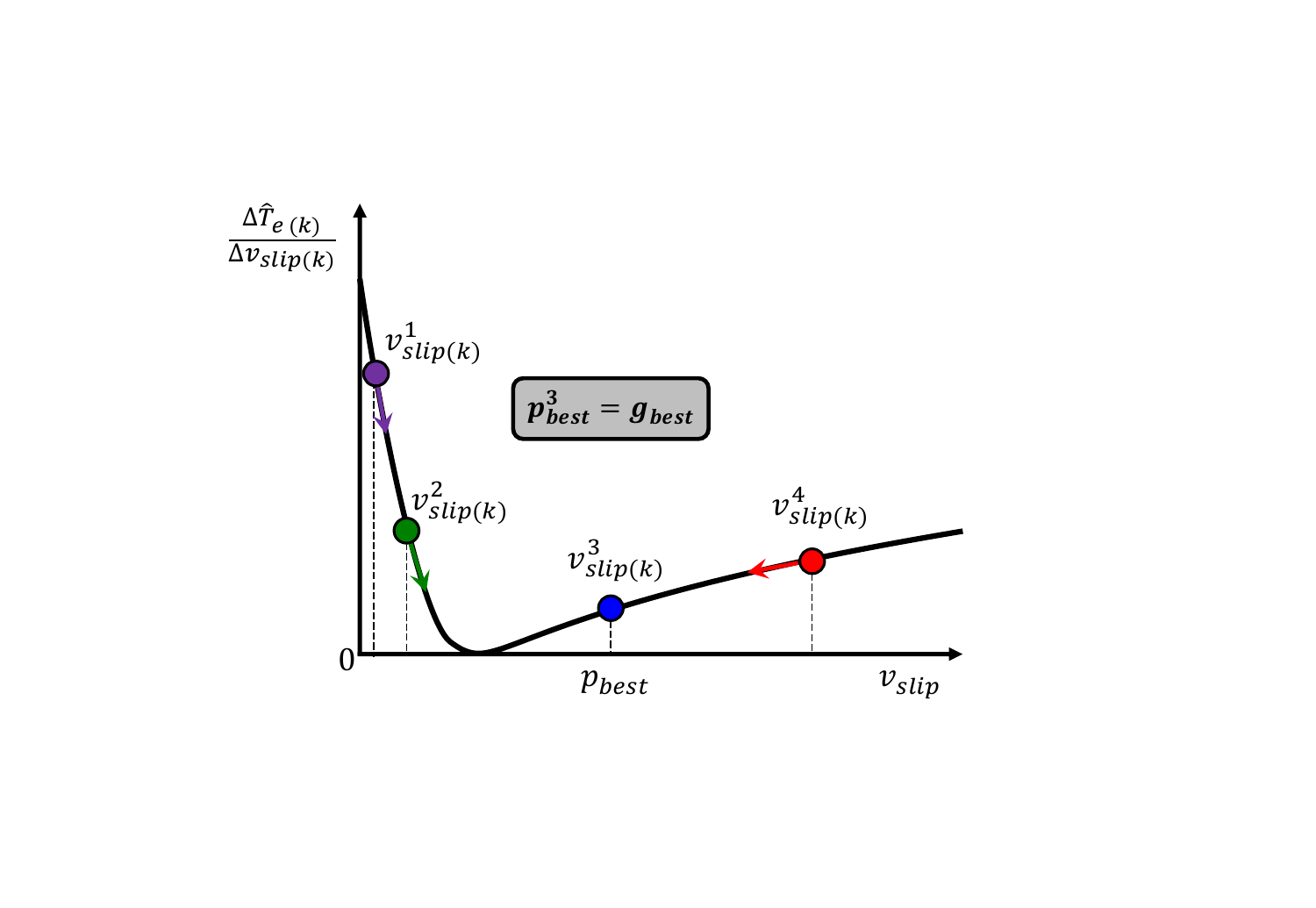}
        \label{fig:pso_1}
    }
     \subfigure[]
    {
        \includegraphics[width=0.4\textwidth,trim={4cm 4cm 4cm 4cmm},clip]{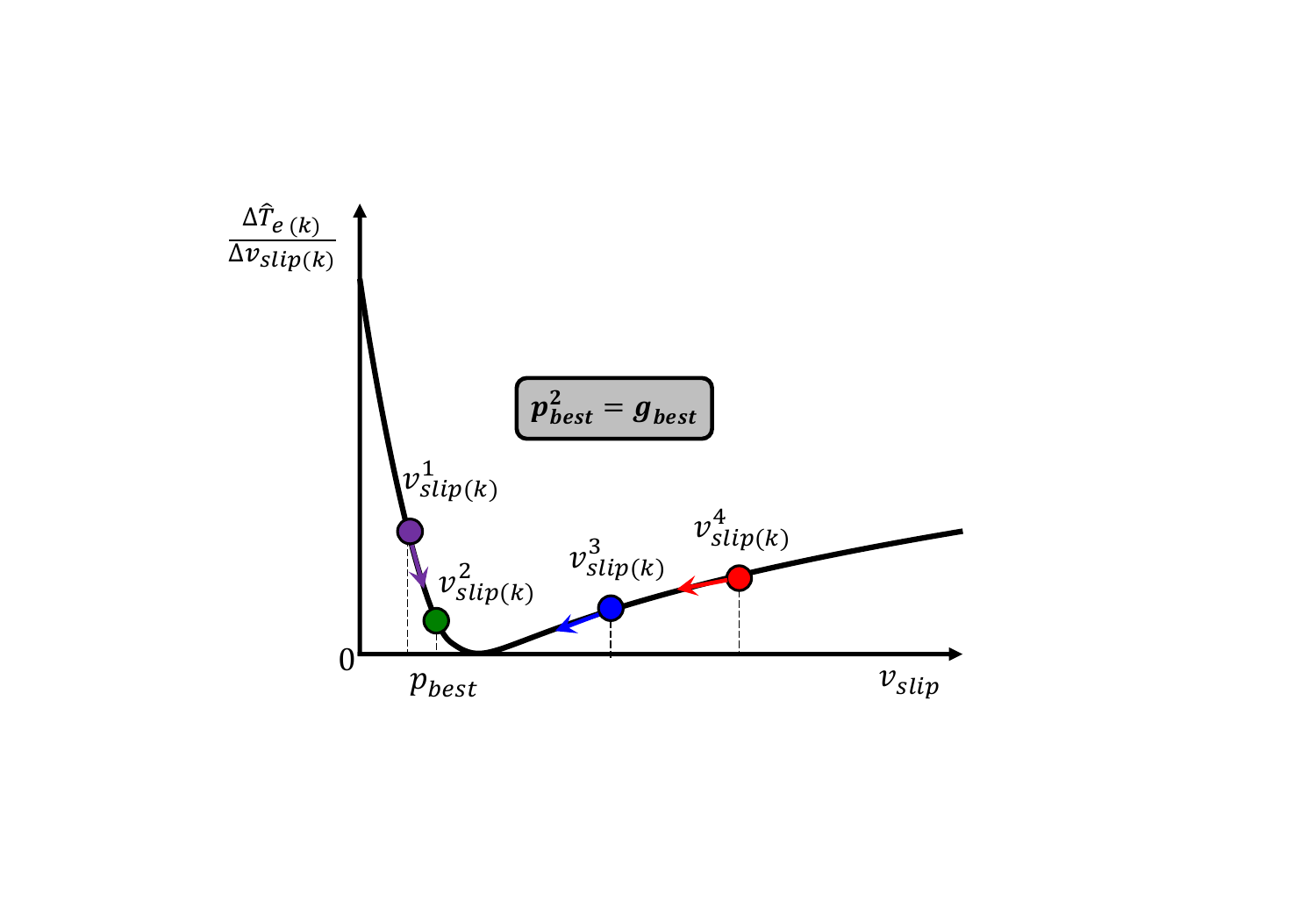}
        \label{fig:pso_2}
    }
         \subfigure[]
    {
        \includegraphics[width=0.4\textwidth,trim={4cm 4cm 4cm 4cmm},clip]{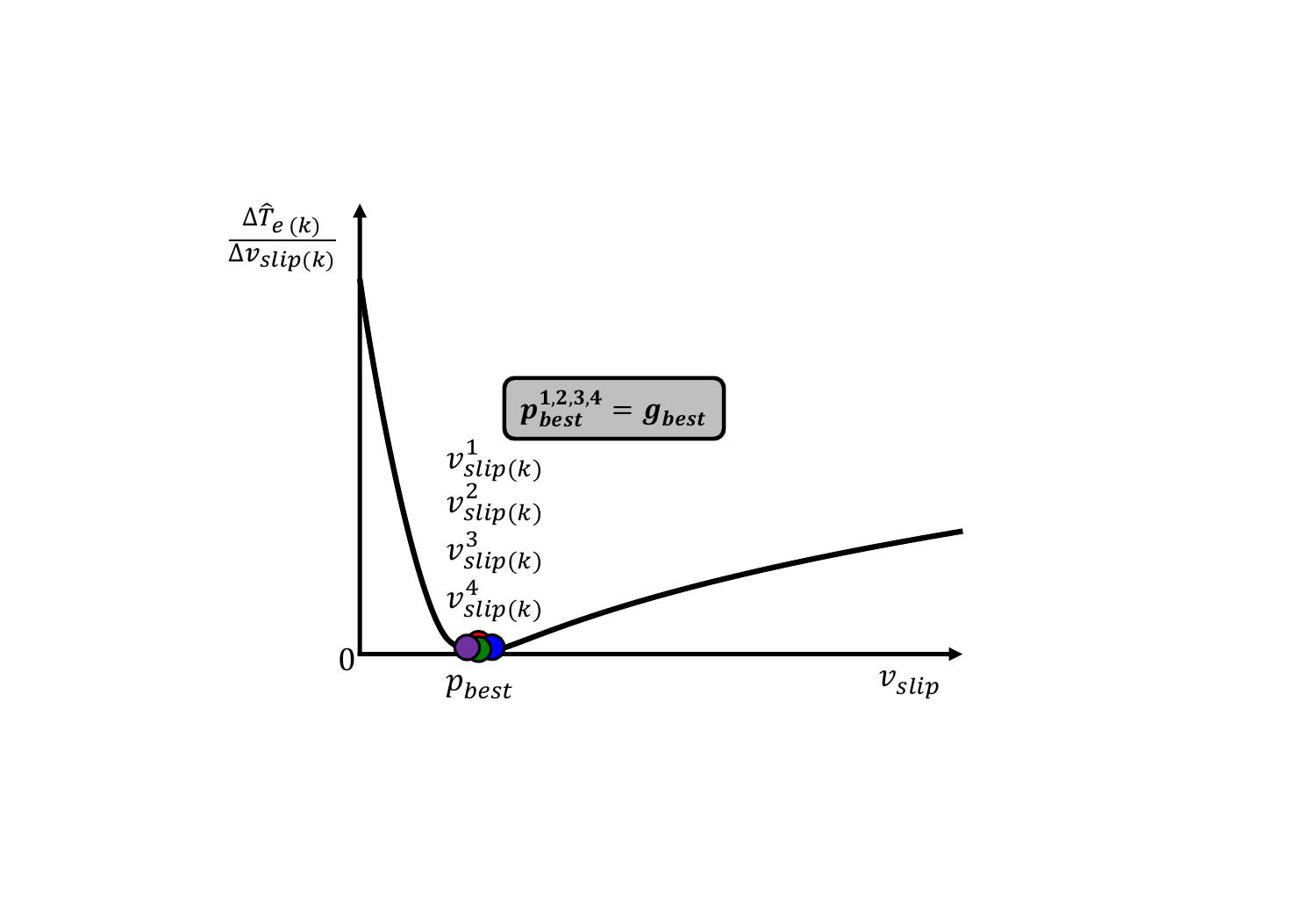}
        \label{fig:pso_3}
    }
        \caption{Procedure of the proposed minimum search using Particle Swarm Optimization (PSO). (a) Particle initialization; (b) Particle movements towards the global best particle after one iteration; (c) Particle swarming towards the global minimum value; (d) Particle final positions at the minimum value where the objective function is achieved.}
    \label{fig:pso_search}
\end{figure}

The PSO algorithm contains a swarm of individuals (particles) at random positions, where each particle represents a possible solution to the problem under investigation. To find the optimal solution, all particles follow a similar behavior, e.g., the position of any particle is influenced by the best particle in the neighborhood ($p_{best}$) as well as the best solution found by all the particles in the entire population ($g_{best}$). The best solution here is referred to the solution which satisfies the selected criterion (fitness function), e.g., to find the global minimum, the global maximum, etc. The particle position adjustment can be represented mathematically as \eqref{eq:x_pso} and \eqref{eq:u_pso}, where $x^{i}_{(k+1)}$ and $u^{i}_{(k+1)}$ represent the current position and velocity of particle $i$ respectively; $w$ is an inertia weighting parameter; $c_1$ and $c_2$ are acceleration coefficients; $r_1$ and $r_2$ are random numbers between 0 to 1; $p^{i}_{best}$ is the best solution of particle $i$ in the previous iteration $k$, $g_{best}$ is the best solution of all particles in the previous iteration $k$.  
\begin{align} 
        u^{i}_{(k+1)} &= w u^{i}_{(k)} + c_1 r_1 (p^{i}_{best}-x^{i}_{(k)}) + c_2 r_2 (g_{best}-x^{i}_{(k)})  \label{eq:u_pso} \\
       x^{i}_{(k+1)} &= x^{i}_{(k)} + u^{i}_{(k+1)} \label{eq:x_pso} 
\end{align}

As in MAT-FLC, the proposed MAT technique using PSO (MAT-PSO) algorithm uses the increments of wheel motor torque $\Delta \hat{T}_{e(k)}$ and slip velocity $\Delta v_{slip(k)}$ to locate the current operating point on the adhesion-slip curve. The output slip velocity reference signal $v^{\ast}_{slip(k)}$ is then adjusted as seen in Fig. \ref{fig:pso_slip}. In the proposed method, the PSO algorithm is designed to search for the minimum absolute value of the adhesion-slip curve slope [see Fig. \ref{fig:pso_search}], as $\frac{\Delta \hat{T}_{e(k)}}{\Delta v_{slip(k)}} \approx 0$ occurs only at the peak of the adhesion-slip curve. 

\begin{figure}[!h]
    \centering
    \includegraphics[width=\columnwidth, trim={0cm 0cm 0cm 0cm},clip]{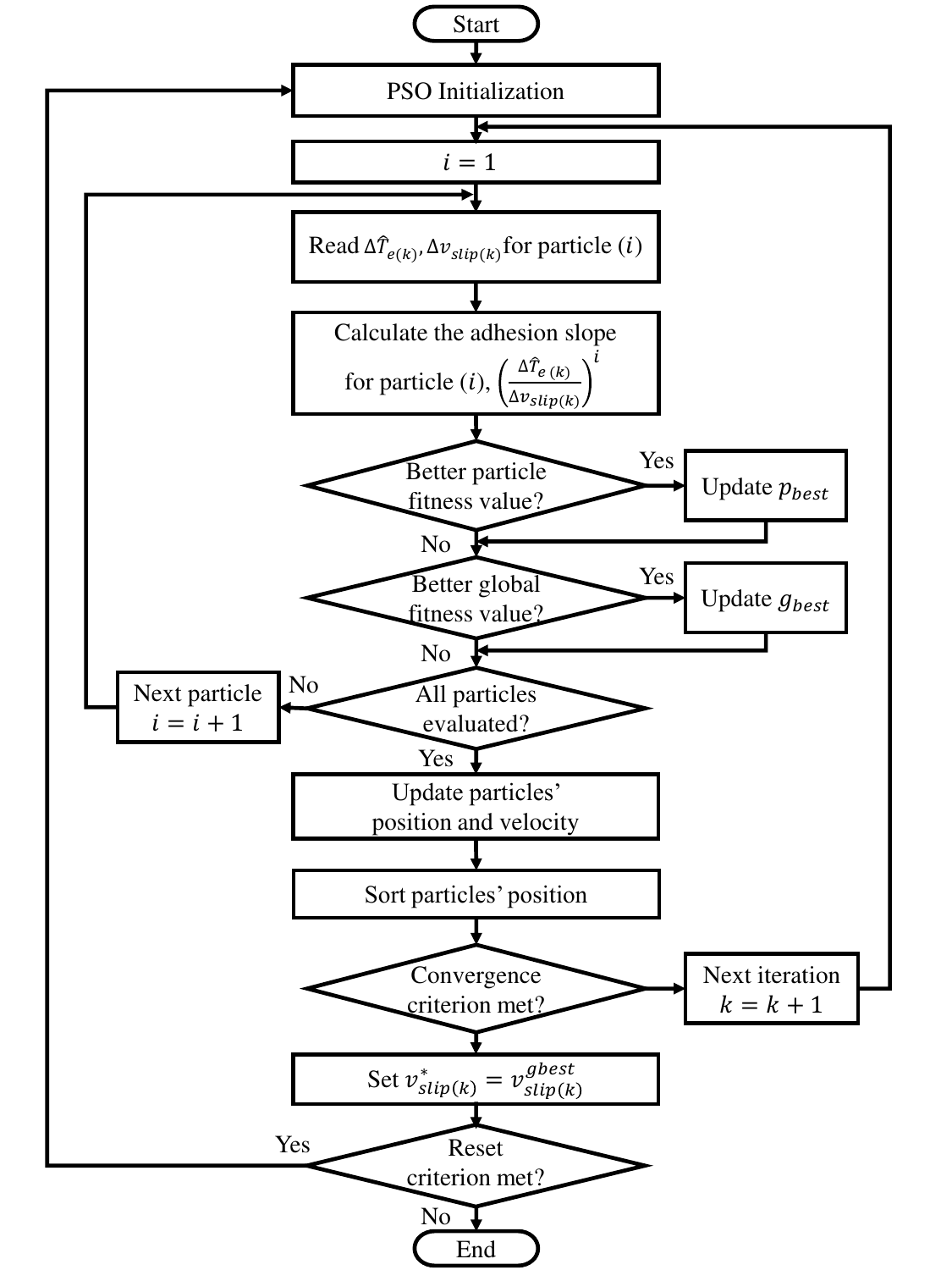}
    \caption{Flowchart of proposed Particle Swarm Optimization (PSO) for minimum search.}
    \label{fig:pso_flowchart}
\end{figure}

The flow chart of the proposed MAT-PSO method is shown in Fig. \ref{fig:pso_flowchart}. Four particles ($N_p=4$) were found adequate to achieve fast search speed with a computational effort suitable for real-time implementation. The algorithm starts with an initial guess of the positions of the particles (i.e. slip velocities) [see Fig. \ref{fig:pso_0}], local best particle position $p_{best}$, and global best particle position $g_{best}$ [see Fig. \ref{fig:pso_1}]. The slope of the adhesion-slip curve is first calculated for each particle and then the fitness function is evaluated individually, where the minimum value is considered to be the local best particle $p_{best}$, and its initial value is updated. 
The new $p_{best}$ value is assigned to be the new global best particle $g_{best}$ which other particles should follow for the next iteration. The $p_{best}$ and $g_{best}$ will be varying while searching for the value that satisfies the fitness function [see particle 2 in Fig. \ref{fig:pso_2}]. Afterward, the output slip velocity $v^{\ast}_{slip(k)}$ is set to be equal to the global best particle position $v^{g_{best}}_{slip(k)}$ [see Fig. \ref{fig:pso_3}]. Finally, the output slip velocity reference $v^{\ast}_{slip(k)}$ will be held constant until the reset function is activated. This occurs when the change in the wheel motor torque and the slip velocity exceeds a certain limit chosen based on the dynamics of the applied system. This situation refers to a change in the adhesion level due to changes in the track condition such as wet, ice, contaminants, etc.

\begin{figure}[t]
    \centering
        \subfigure[]
    {
            \includegraphics[width=0.8\columnwidth,trim={0cm 0cm 0cm 0cm},clip]{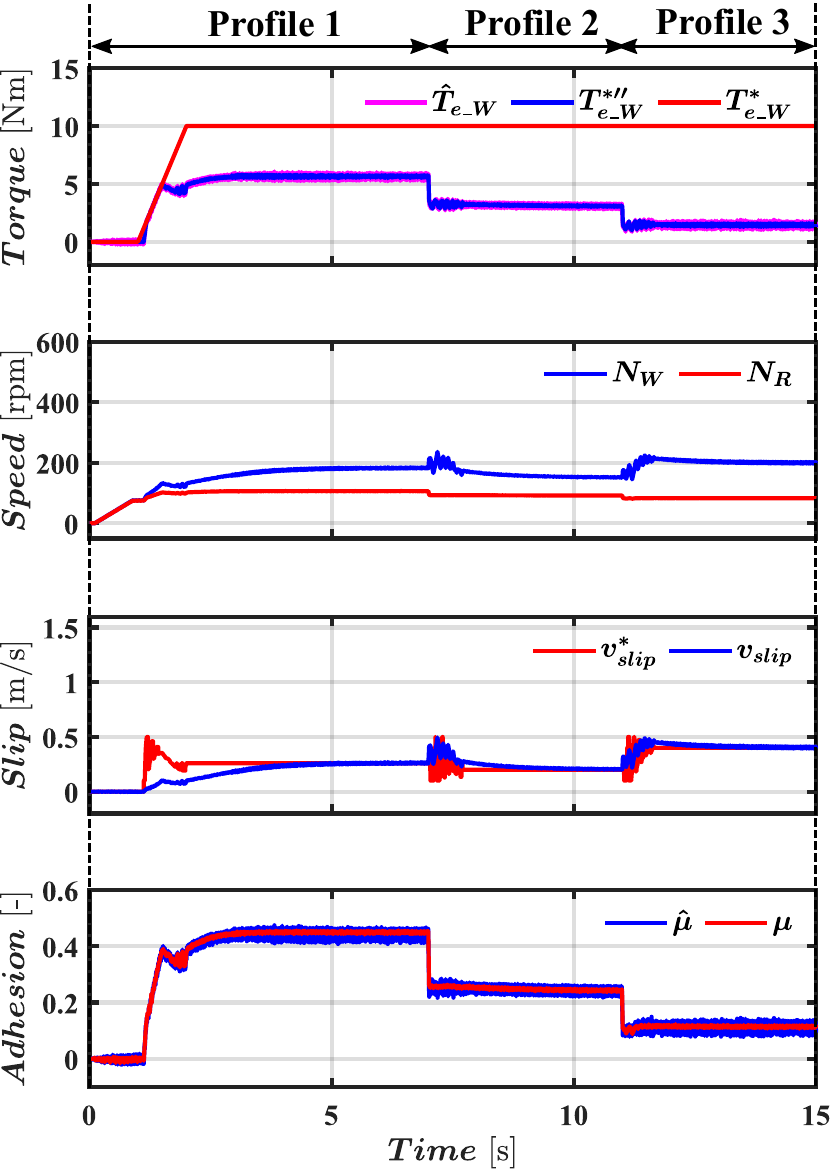}
        \label{fig:time_pso}
    }
    \subfigure[]
    {
    \includegraphics[width=0.85\columnwidth, trim={0cm 0cm 0cm 0cm},clip]{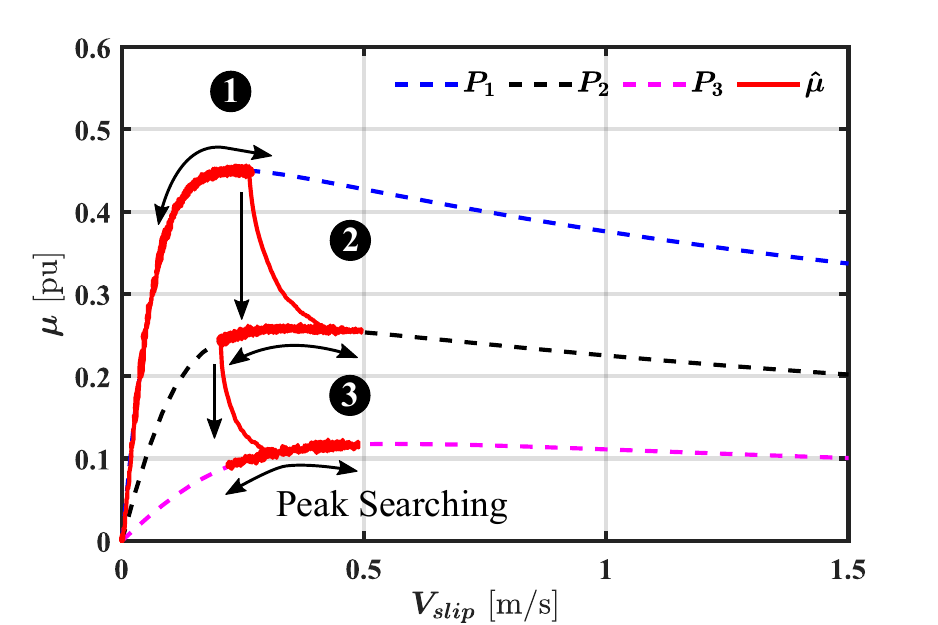}
    \label{fig:mu_pso}
    } 
    \caption{Simulation results: Proposed MAT using Particle Swarm Optimization (PSO) (MAT-PSO).}
    \label{fig:result_pso}
\end{figure}
    For practical purposes and safety concerns in railways, it is important to note that the particle inertia weighting parameter $w$ and the acceleration coefficients $r_1$ and $r_2$ in \eqref{eq:u_pso} should be $\leq$ 1.0. Four particles have been chosen as they allow to cover a wide range of slip velocities in the adhesion-slip characteristics curve with reduced search time.  
    
From Fig. \ref{fig:result_flc} and Fig. \ref{fig:result_pso}, it is observed that MAT-PSO has a slightly slower response compared to MAT-FLC. This is due to the re-initialization of particles' positions and random movement when the reset function is activated [see third subplot in Fig. \ref{fig:time_pso}]. On the other hand, PSO algorithm shows a superior steady-state performance for obtaining the correct slip velocity command value at which the peak of the adhesion curve occurs. This can be noticed for $P_3$ where it reaches to $v^{\ast}_{slip}=0.41$ m/s while the theoretical peak occurs at $v^{\ast}_{slip}=0.45$ m/s [see Fig. \ref{fig:mu_pso}].

\begin{figure}[!h]
    \centering
    \includegraphics[width=\columnwidth, trim={4cm 5cm 4cm 2.5cm},clip]{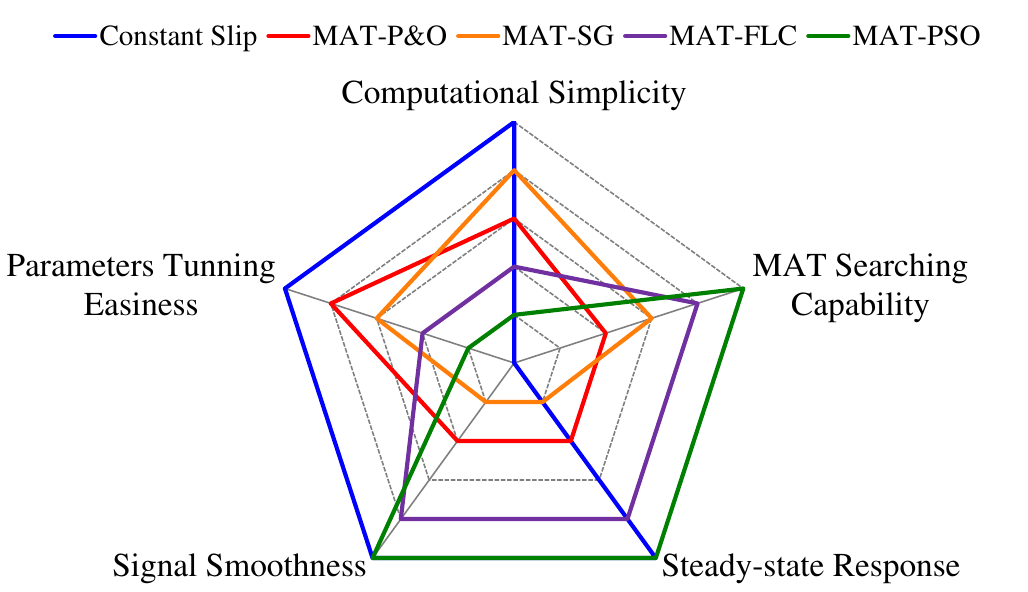}
    \caption{Comparison of slip velocity control methods.}
    \label{fig:comparison}
\end{figure}

Fig. \ref{fig:comparison} summarizes the main characteristics and expected performance of the methods being considered. Constant slip velocity control is seen to provide excellent results in almost all the aspects being evaluated, but this is at the price of no MAT searching capability. This control mode would be beneficial for rail tracks with known adhesion characteristics. Unfortunately, this knowledge is not available in practice. The shortcomings of the constant slip method are overcome using the proposed MAT-PSO, but at the cost of implementation complexity and difficult tuning. The proposed MAT-FLC shows a moderate performance, providing some of the advantages of MAT-PSO but with less computational effort. MAT-P\&O and MAT-SG show similar performance regarding tracking capability and simplicity. However, MAT-SG shows the worst performance regarding signal smoothness and steady-state response.
\section{Experimental Validation} \label{sec:Exp}

In this section, all methods discussed in section \ref{sec:overview_slip_control} and the proposed ones in section \ref{sec:proposed} are validated and evaluated experimentally. 

\subsection{Test Bench Setup} \label{subsec:testbench}

\begin{figure*}[!ht]
    \centering
    \subfigure[]
    {
        \includegraphics[width=0.5\textwidth,trim={0cm 0cm 0cm 0cm},clip]{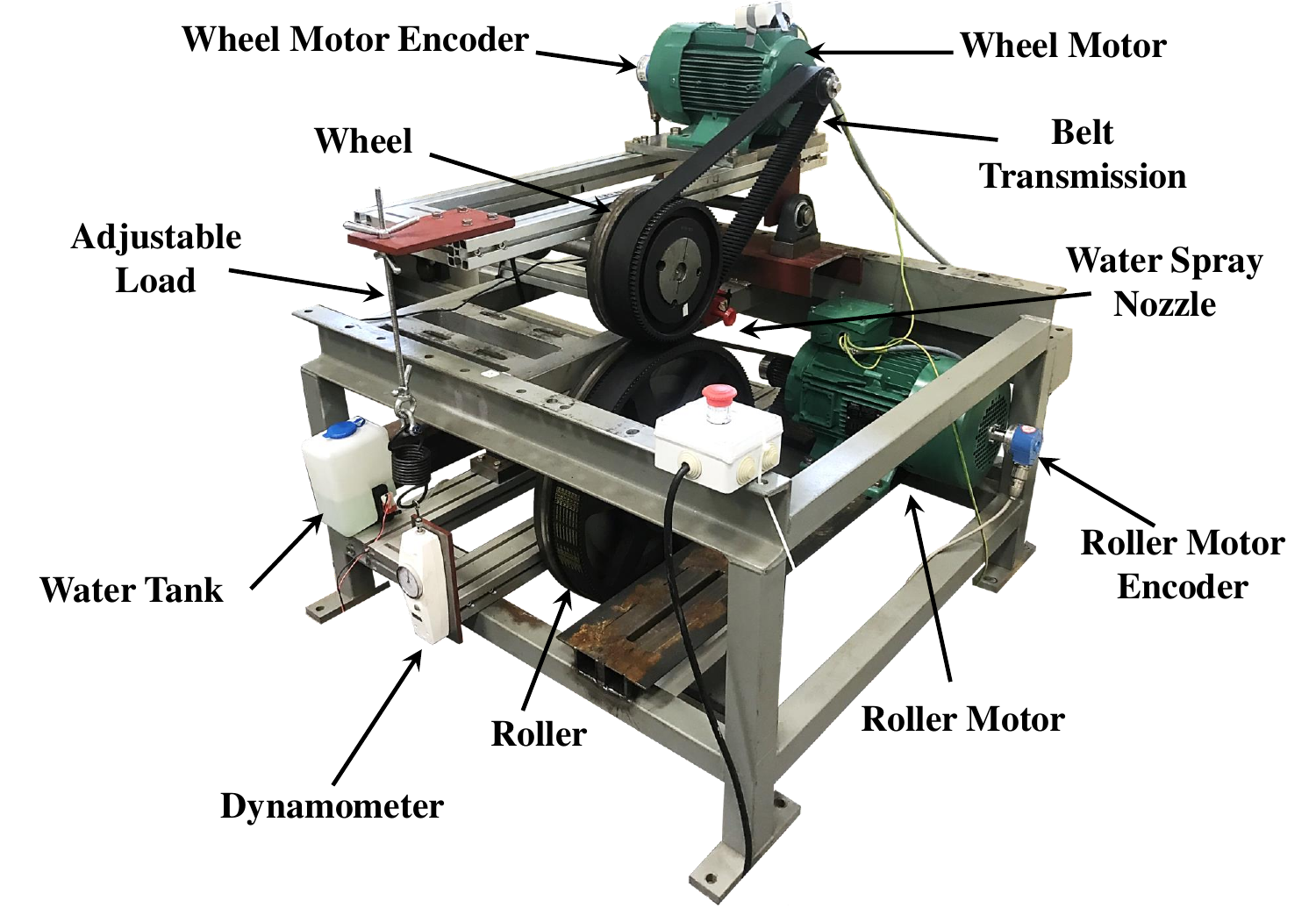}
        \label{fig:tesebench_rollerrig}
    }
    \\
    \subfigure[]
    {
        \includegraphics[width=0.4\textwidth,trim={0 0cm 0 0cm},clip]{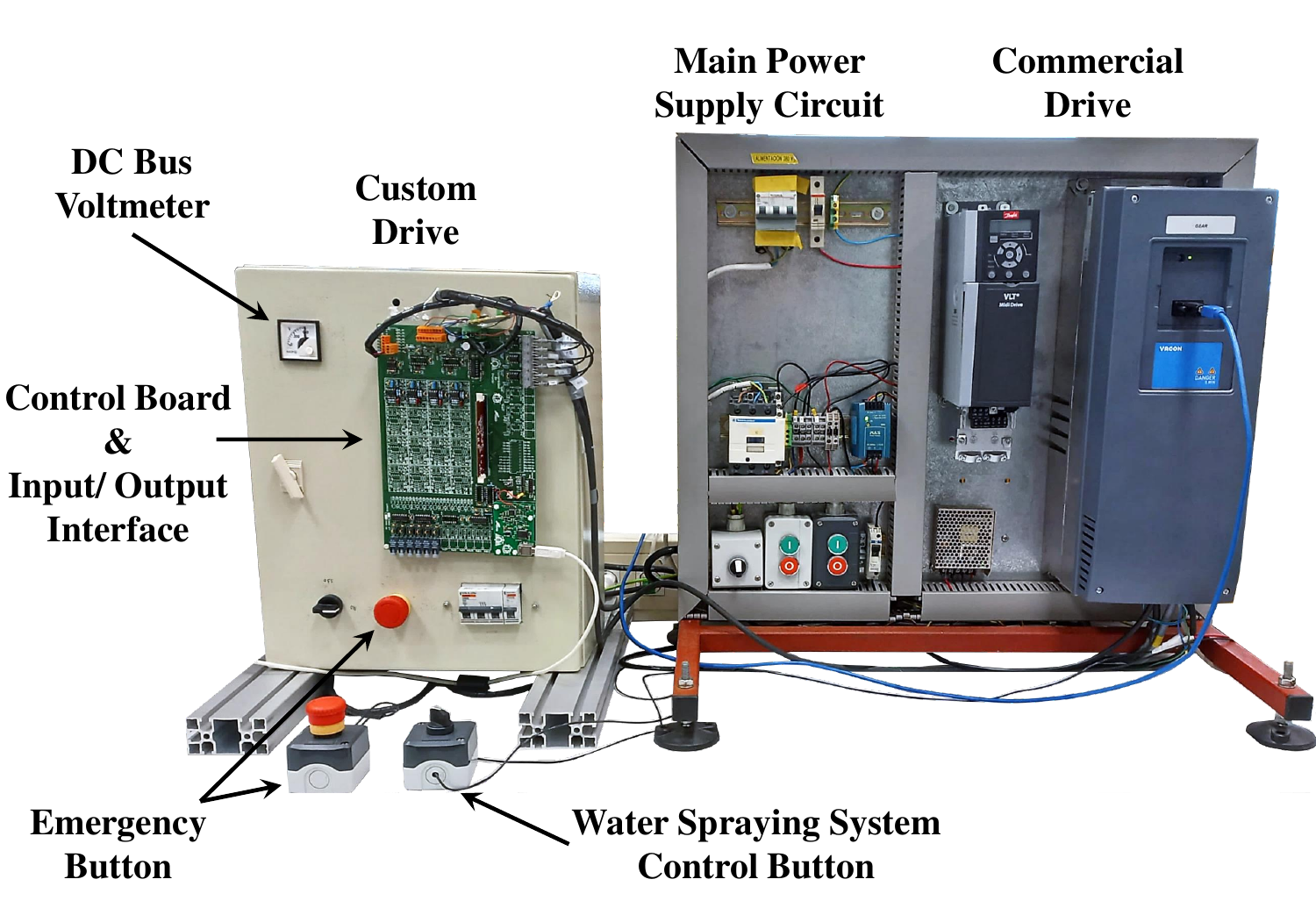}
        \label{fig:tesebench_drive}
    }
        \subfigure[]
    {
        \includegraphics[width=0.4\textwidth,trim={0 0cm 0 0cm},clip]{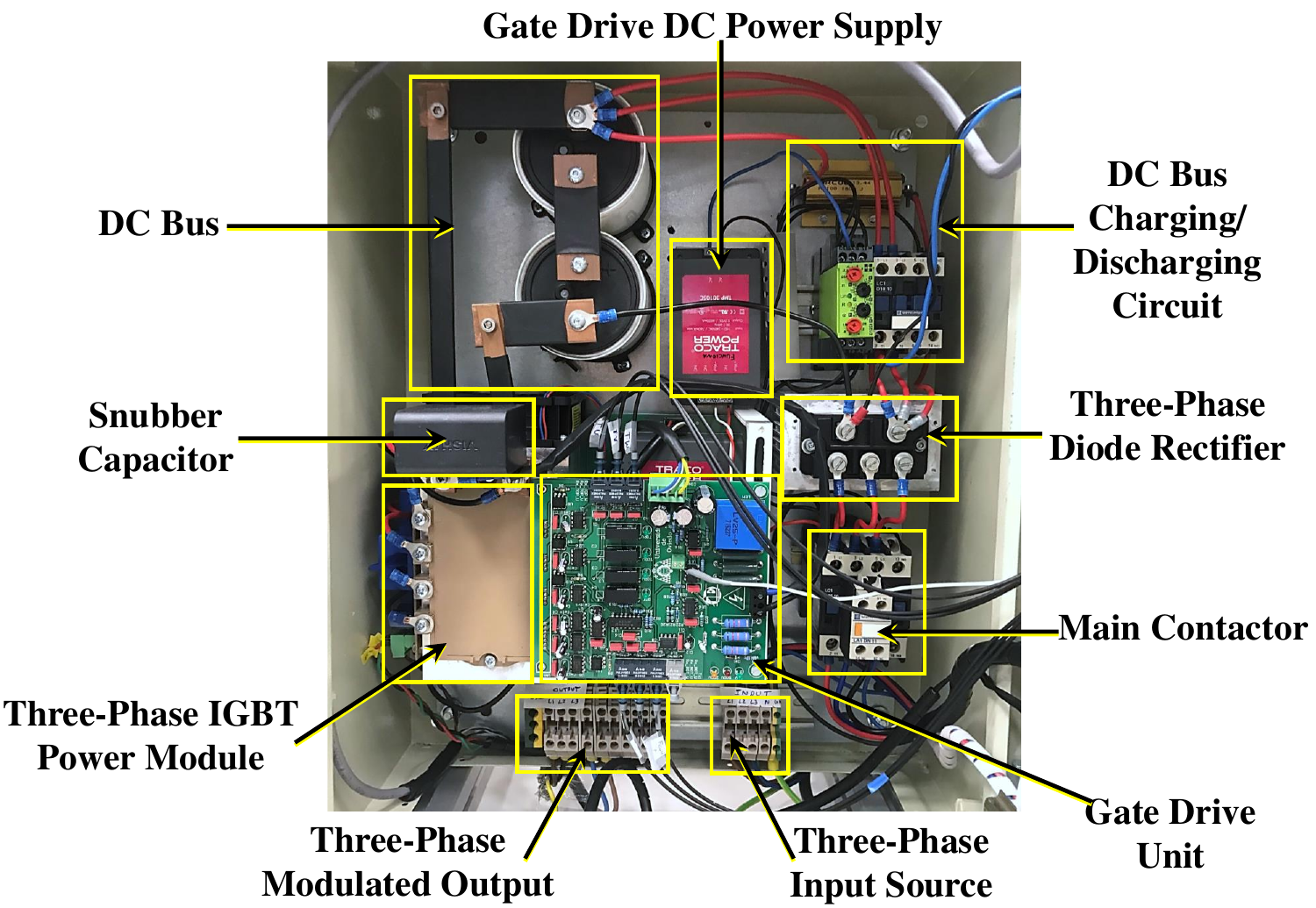}
        \label{fig:tesebench_custom_drive}
    }
        \caption{Overview of the experimental setup: (a) roller rig test bench; (b) Electrical circuit and motor drives; (c) Custom drive elements.}
    \label{fig:tesbench}
\end{figure*}

The proposed scaled roller rig described in subsection \ref{subsec:scaled_roller_rig} is used for emulating the wheel-rail contact dynamics. The roller rig test bench is shown in Fig. \ref{fig:tesebench_rollerrig} and its parameters are given in table \ref{Table 1} where inertia is calculated in wheel and roller motor reference frame respectively \cite{nihal2022design}. The roller rig load is adjusted by a spring system attached to a dynamometer for load force measurements which applies extra force on the wheel as seen in Fig. \ref{fig:tesebench_rollerrig}. Additionally, a water spraying system is included for evaluation of the control strategies when the wheel-roller surface becomes wet. 

\begin{table}[ht]
\caption{Scaled roller rig parameters}
\label{Table 1}
\resizebox{\columnwidth}{!}{%
\begin{tabular}{ccccc}
\hline \hline
System                            & Parameter          & Wheel & Roller & Unit    \\ \hline 
\multirow{3}{*}{Wheel and Roller} & Radius             & 0.125 & 0.25   & m       \\ 
                                  & Force              & 843   & 843    & N       \\ 
                                  & Torque             & 105.3 & 210.7  & Nm      \\ \hline
Transmission                      & Gear ratio         & 90/24 & 192/26 & -       \\ \hline
\multirow{5}{*}{Traction Motor}   & Rating             & 4     & 5.5    & kW      \\ 
                                  & Power              & 1.78  & 1.78   & kW      \\ 
                                  & Torque             & 28.1  & 28.5   & Nm      \\ 
                                  & Speed              & 604.8 & 595.5  & rpm     \\  
                                  & Encoder Resolution & 500   & 500    & ppr     \\ \hline
Inverter                          & Rating             & 4     & 15     & kW      \\ \hline
Motor-Wheel                       & Inertia            & 0.002 & 0.007  & kgm$^2$ \\ \hline \hline
\end{tabular}%
}
\end{table}

Two four-pole induction motors of 4 kW and 5.5 kW are used for wheel and roller respectively (see Fig. \ref{fig:tesebench_rollerrig}). The induction motors are driven by two drives, both using rotor field-oriented control (RFOC) (see Fig. \ref{fig:tesebench_drive}) \cite{blaschke1972principle, zhou2002relationship}. A commercial VACON NXP00385 Danfoss drive is used in speed control mode to keep the roller motor speed constant. A custom drive is used to feed the wheel motor. All the methods discussed in this paper are implemented in a TMDSCNCD28335 digital signal processor used to control the custom drive \ref{sec:overview_slip_control} and \ref{sec:proposed}. 

\subsection{Methodology} \label{subsec:methedology}

The methods discussed in Sections \ref{sec:overview_slip_control} and \ref{sec:proposed} were tested and validated in the test bench. For each method, the experiments were realized under dry and wet operating conditions. This is the normal procedure in field tests. Wet tests were done by spraying water with a flow rate of 140 ml/min. The experiments were conducted at a room temperature of 20-25 $^\circ$C.

For each test, the following sequence is followed:

\begin{enumerate}
    
    \item Experiments always start with a dry wheel-roller contact point. The roller speed controller is enabled and set to 60 rpm and is kept constant during the rest of the experiment. This would correspond to the train running freely. This assumption is valid for the case of very high inertia systems such as a train. 

    \item At $t\approx5$ s the wheel torque controller is activated and a torque of 10 Nm is commanded; the torque command gradient is set to 5 Nm/s. The wheel will start slipping if the commanded torque surpasses the adhesion level, which will depend on the wheel-roller contact condition. If this happens, the slip velocity controller will be activated either for limiting slip velocity or controlling it for tracking the maximum adhesion level.

    \item At $t\approx35$ s the water spraying system is turned on, the wheel-roller adhesion reducing. 

    \item At $t\approx65$ s the wheel torque command is set to zero till the end of the experiment. 

\end{enumerate}

The measured data is post-processed using MATLAB to obtain the traction motor torque $\hat{T}_{e\_W}$, wheel-roller slip velocity between the wheel and the roller $V_{slip}$, and estimated adhesion coefficient $\mu$.
\subsection{Results and Discussion} \label{subsec:esp_results}
The experimental measurements are used to validate the performance of the proposed MAT algorithms. Also, to assess the effect of changing adhesion conditions, on the overall performance of the addressed maximum adhesion tracking strategies. Finally, a comparative analysis is carried out between the existing and proposed MAT methods. 

\begin{figure*}[t]
    \centering
        \subfigure[]
    {
            \includegraphics[width=0.31\textwidth,trim={0cm 0cm 0cm 0cm},clip]{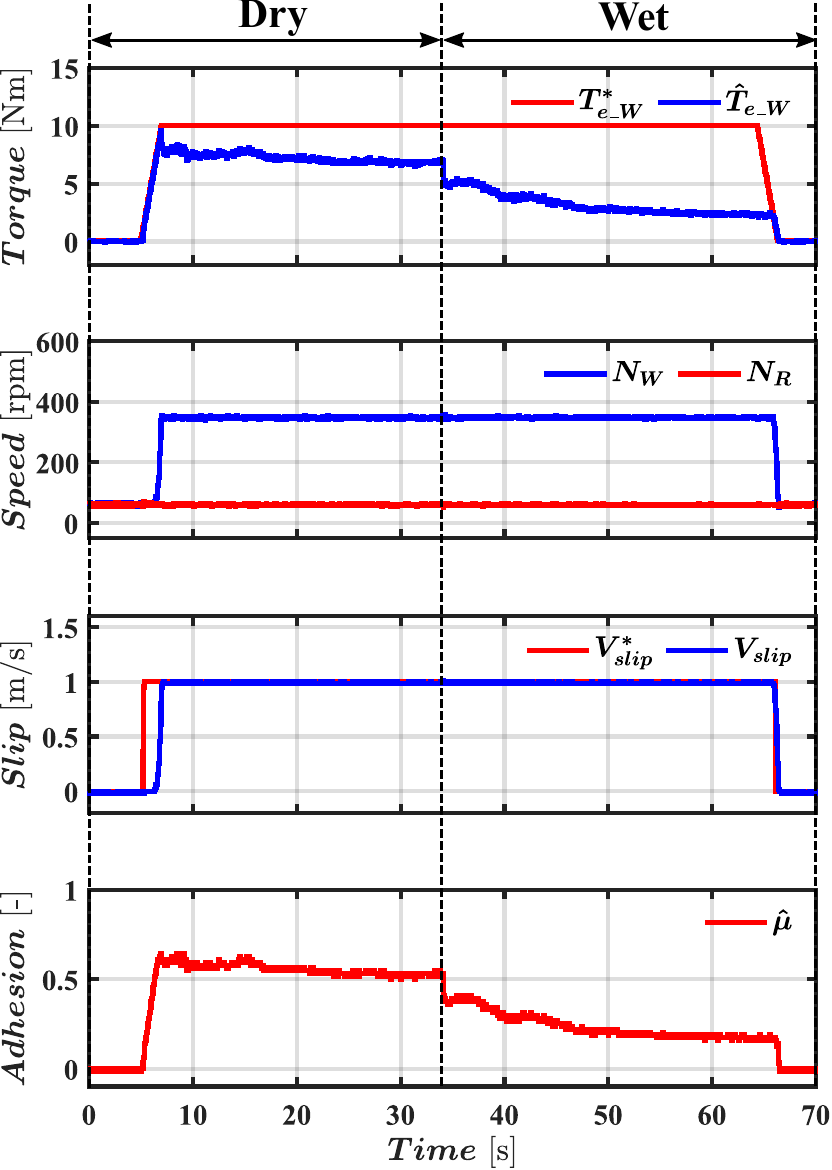}
        \label{fig:exp_time_const}
    }
                \subfigure[]
    {
            \includegraphics[width=0.31\textwidth,trim={0cm 0cm 0cm 0cm},clip]{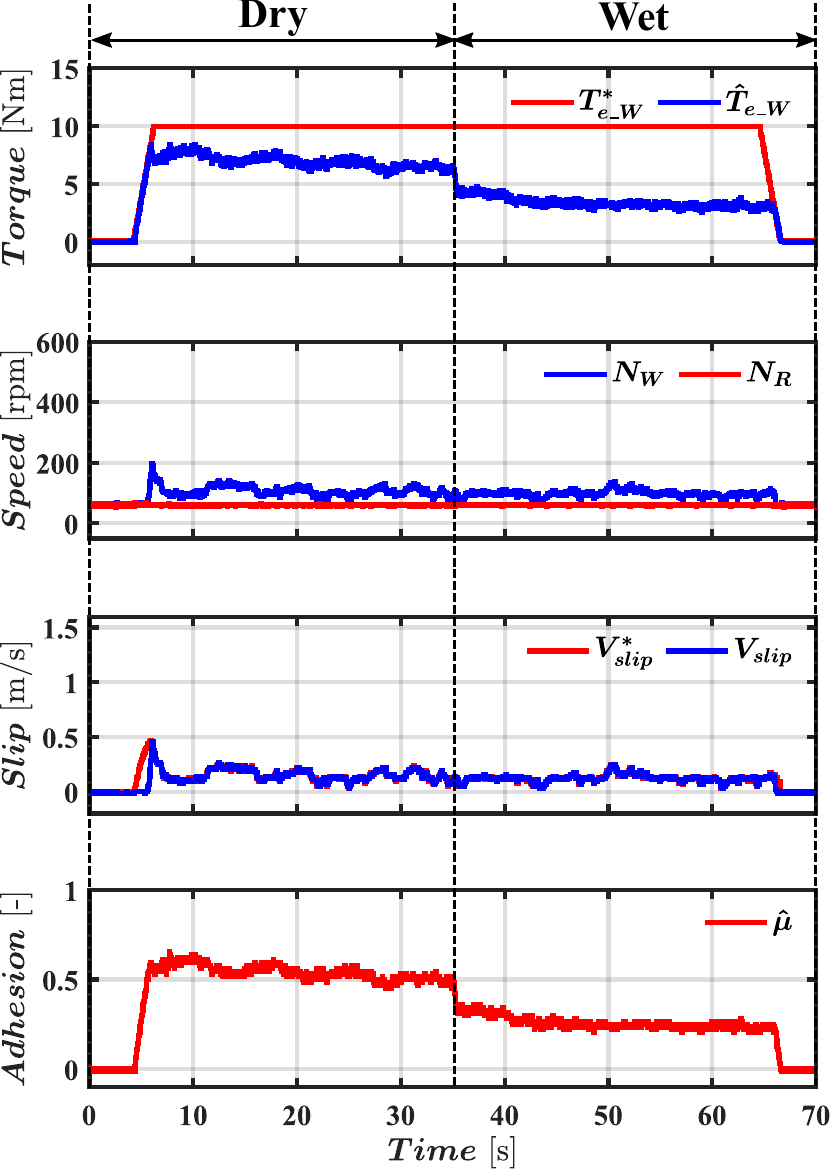}
        \label{fig:exp_time_po}
    }
                \subfigure[]
    {
            \includegraphics[width=0.31\textwidth,trim={0cm 0cm 0cm 0cm},clip]{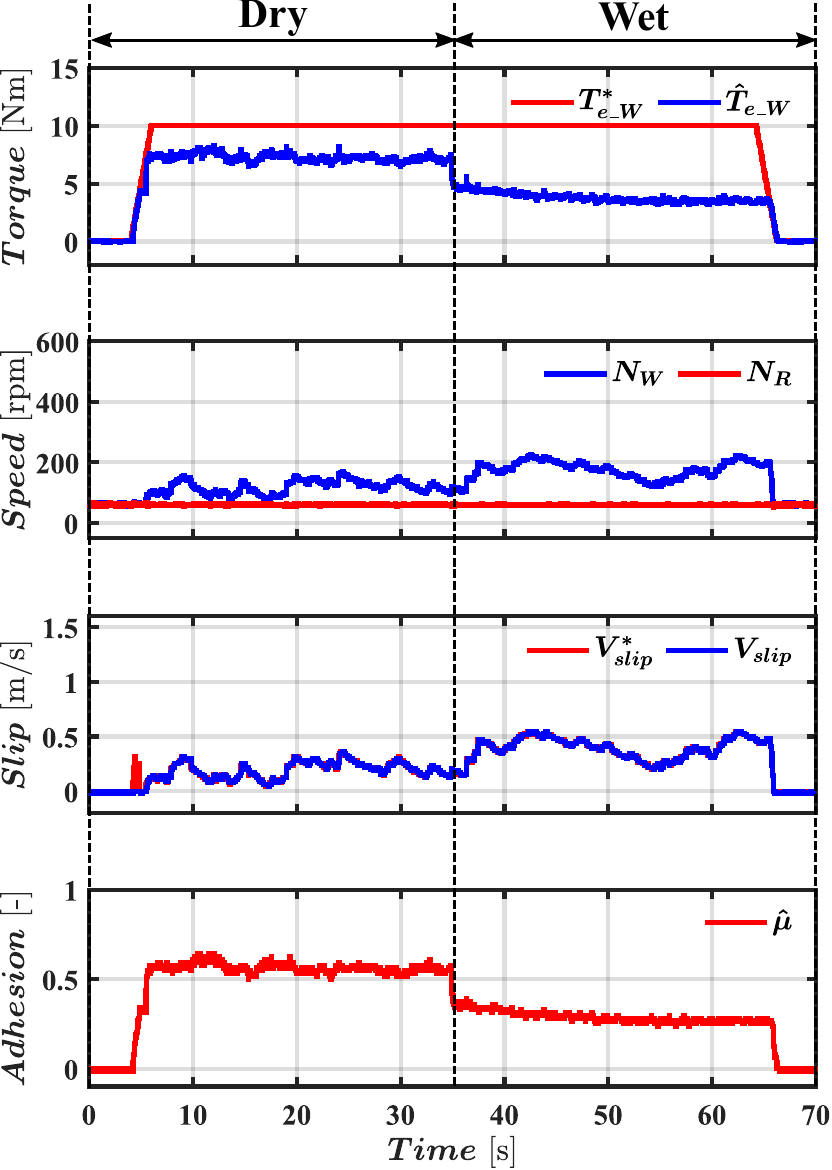}
        \label{fig:exp_time_steepest}
    }
          \hfill
                \subfigure[]
    {
            \includegraphics[width=0.31\textwidth,trim={0cm 0cm 0cm 0cm},clip]{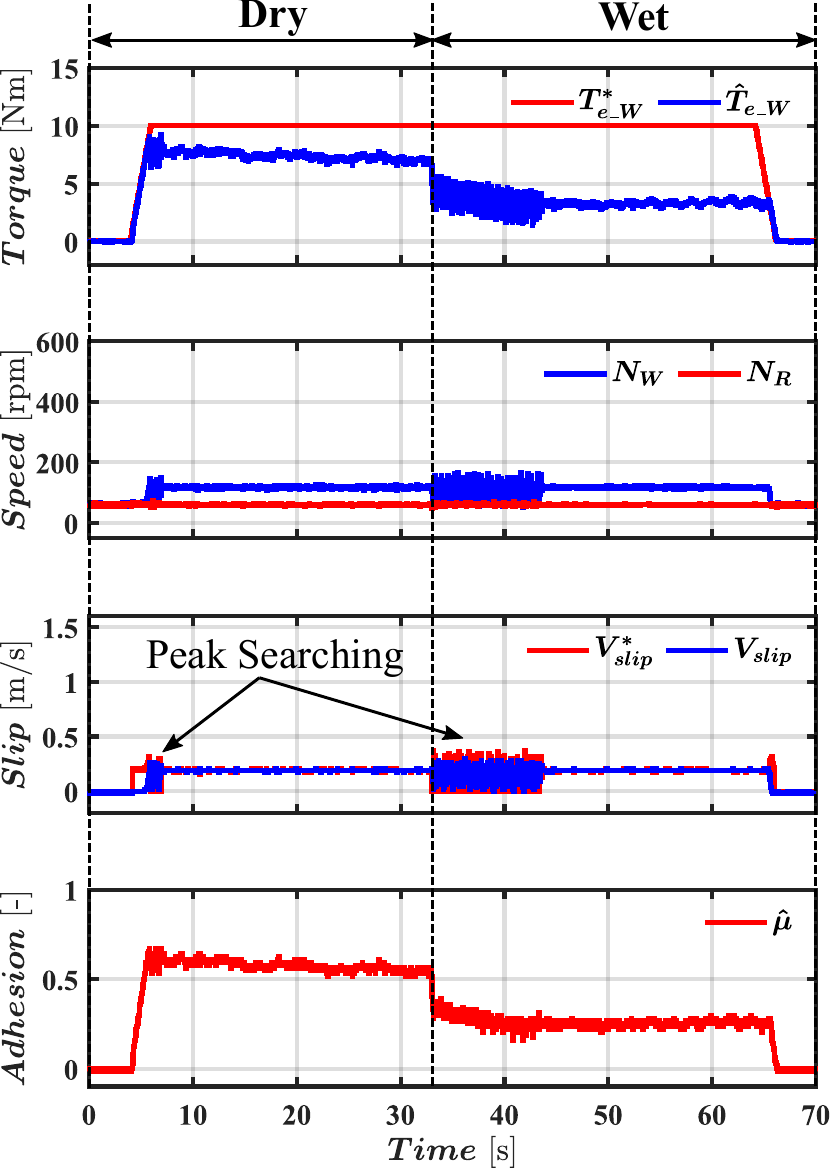}
        \label{fig:exp_time_flc}
    }
                \subfigure[]
    {
            \includegraphics[width=0.31\textwidth,trim={0cm 0cm 0cm 0cm},clip]{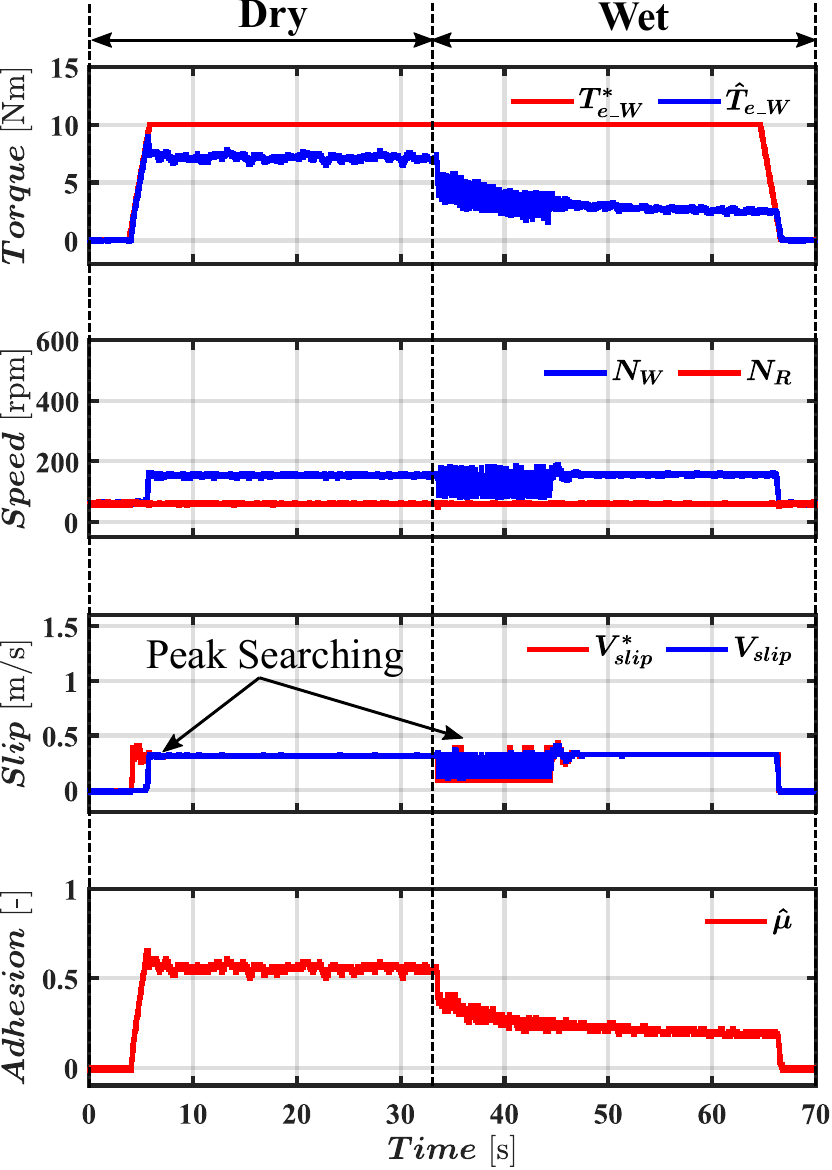}
        \label{fig:exp_time_pso}
    }      
    
    \caption{Experimental results. Response in the time domain: (a) constant slip velocity; (b) MAT-P\&O; (c) MAT-SG; (d) Proposed MAT-FLC; and (e) Proposed MAT-PSO. }
    \label{fig:exp_result_time}
\end{figure*}

Overall, experimental results for all methods are in good agreement with the simulation results obtained in Sections \ref{sec:overview_slip_control} and \ref{sec:proposed}. However, torque and speed oscillations in the experiments are seen to be lower than in simulation (see Fig. \ref{fig:time_po} \& \ref{fig:time_steepest} vs. \ref{fig:exp_time_po} \& \ref{fig:exp_time_steepest}). Contrary to simulation results, changes from dry to wet conditions of the contact point do not occur instantaneously in the test bench. This would explain some of the differences observed between simulation and experimental results. 

\begin{figure*}[t]
    \centering
        \subfigure[]
    {
            \includegraphics[width=0.31\textwidth,trim={0cm 0cm 0cm 0cm},clip]{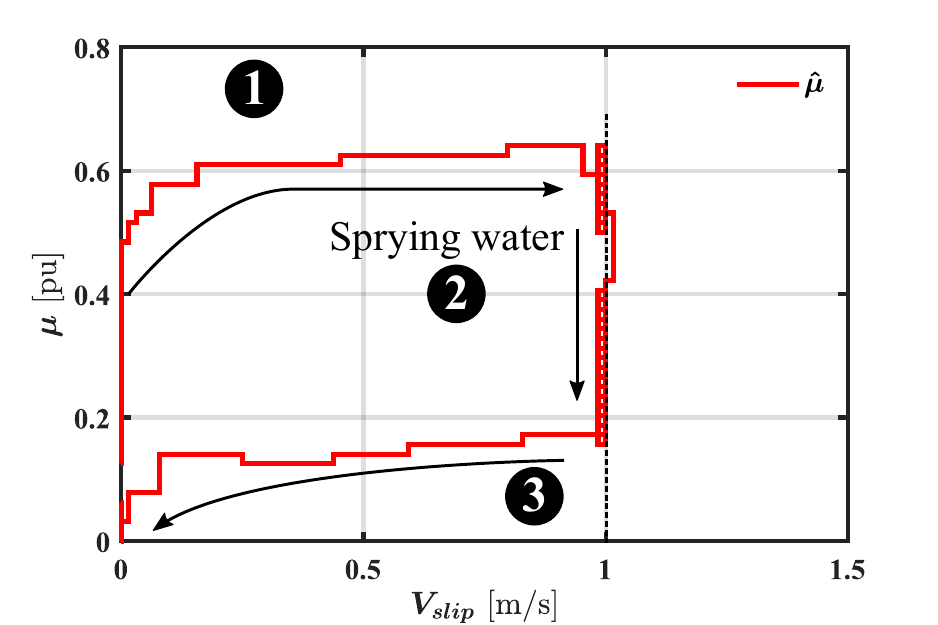}
        \label{fig:exp_mu_const}
    }
                \subfigure[]
    {
            \includegraphics[width=0.31\textwidth,trim={0cm 0cm 0cm 0cm},clip]{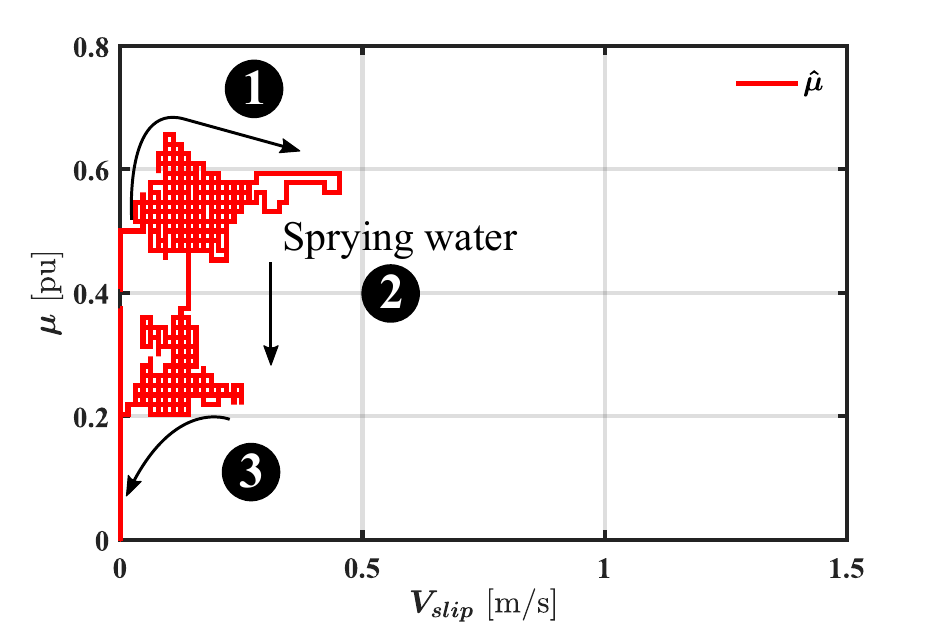}
        \label{fig:exp_mu_po}
    }
                \subfigure[]
    {
            \includegraphics[width=0.31\textwidth,trim={0cm 0cm 0cm 0cm},clip]{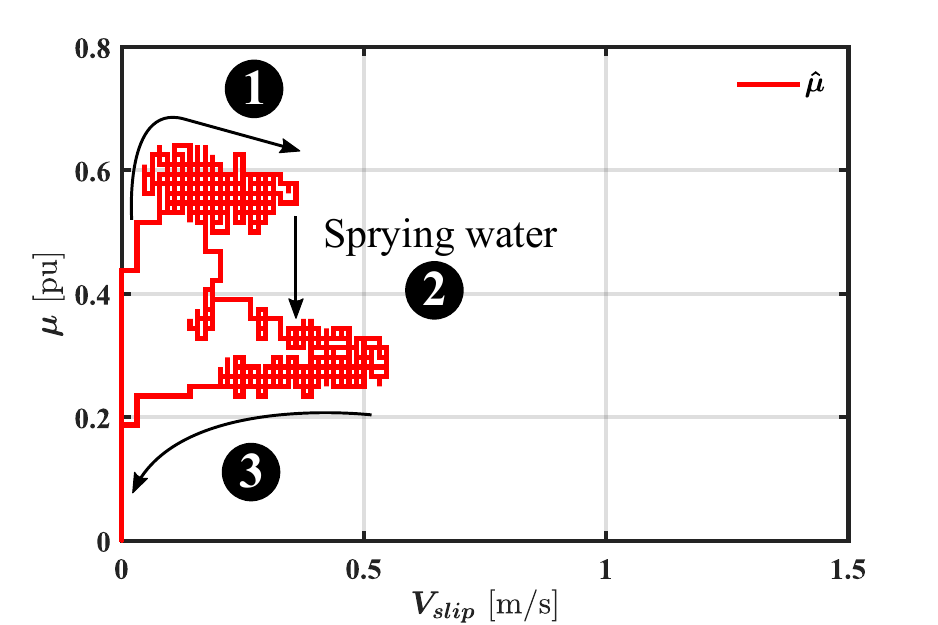}
        \label{fig:exp_mu_steepest}
    }
                \subfigure[]
    {
            \includegraphics[width=0.31\textwidth,trim={0cm 0cm 0cm 0cm},clip]{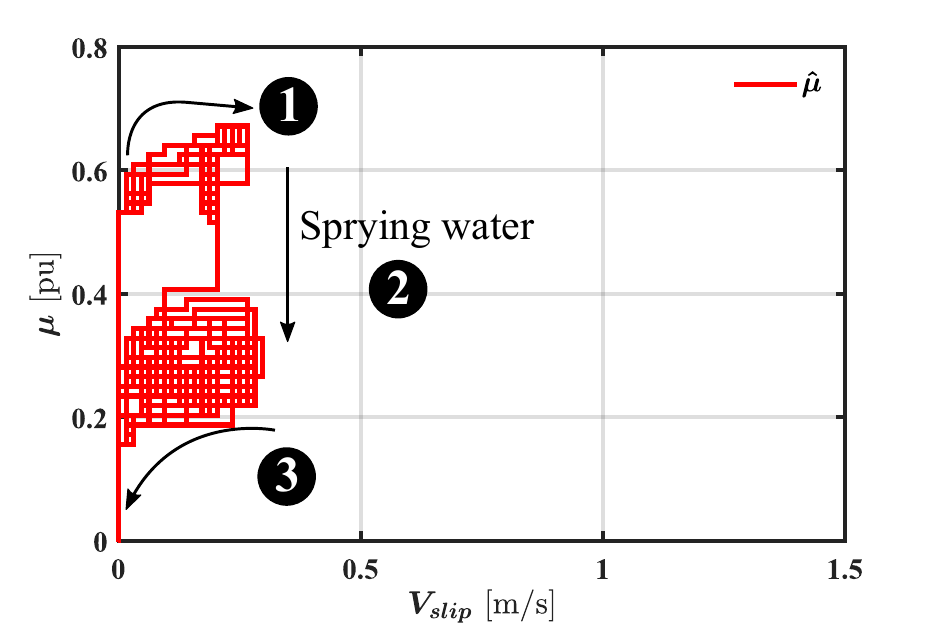}
        \label{fig:exp_mu_flc}
    }
                \subfigure[]
    {
            \includegraphics[width=0.31\textwidth,trim={0cm 0cm 0cm 0cm},clip]{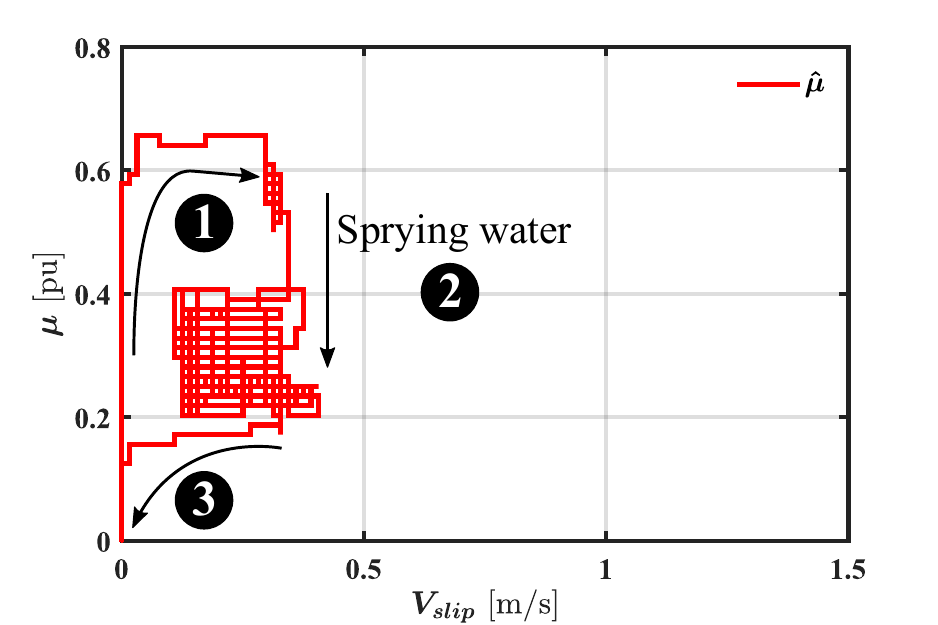}
        \label{fig:exp_mu_pso}
    }      
    
    \caption{Experimental results. Adhesion-slip trajectory: (a) constant slip velocity; (b) MAT-P\&O; (c) MAT-SG; (d) Proposed MAT-FLC; and (e) Proposed MAT-PSO. }
    \label{fig:exp_result_mu}
\end{figure*}

For the constant slip control method (see Fig. \ref{fig:exp_time_const} and \ref{fig:exp_mu_const}), the achieved steady-state adhesion coefficient for dry and wet conditions are $\hat{\mu}\approx 0.5$ and $\hat{\mu}\approx 0.15$ respectively. These values increase to $\hat{\mu}\approx 0.6$ and $\hat{\mu}\approx0.25$ for all MAT methods considered in this paper (see Fig. \ref{fig:exp_time_po} to Fig. \ref{fig:exp_time_pso} and Fig. \ref{fig:exp_mu_po} to Fig. \ref{fig:exp_mu_pso}). As expected, all the methods track the peak of the adhesion curve both in dry and wet conditions. An increase in slip velocity ($v_{slip}\approx0.45$ m/s) is noticed while peak searching in wet condition (see Fig. \ref{fig:exp_time_steepest} $t=45$ s to $t=65$ s) with MAT-SG due to the continuous increment of ($\Delta v^{\ast}{slip(k)}$) term in the slip velocity command in the case of flat adhesion curves aiming to find the maximum peak. This doesn't occur with MAT-PO as the search logic alternate between the increment/decrement of the slip velocity command. 

The two proposed methods MAT-FLC and MAT-PSO show a similar response as seen in Fig. \ref{fig:exp_time_flc} and Fig. \ref{fig:exp_time_pso}, as well as Fig. \ref{fig:exp_mu_flc} and \ref{fig:exp_mu_pso}. However, for MAT-FLC the slip velocity command is kept at the same value ($v_{slip}\approx0.2$ m/s) for both adhesion conditions while for MAT-PSO slip velocity is slightly differs as ($v_{slip}\approx0.3$ m/s) and ($v_{slip}\approx0.35$ m/s) in dry and wet conditions respectively. The difference in the performance of FLC and PSO algorithms is negligible and can be only noticed in the search space of the adhesion-slip curves, but noting the relevant differences in the implementation complexity. Therefore, the proposed MAT methods achieve the same adhesion level ($\approx 20\%$ higher than the constant slip method) with less slip velocity in steady-state compared to existing methods. 

\begin{figure*}[!h]
    \centering
    \subfigure[]
    {
        \includegraphics[width=0.45\textwidth,trim={4cm 3cm 3cm 4cm},clip]{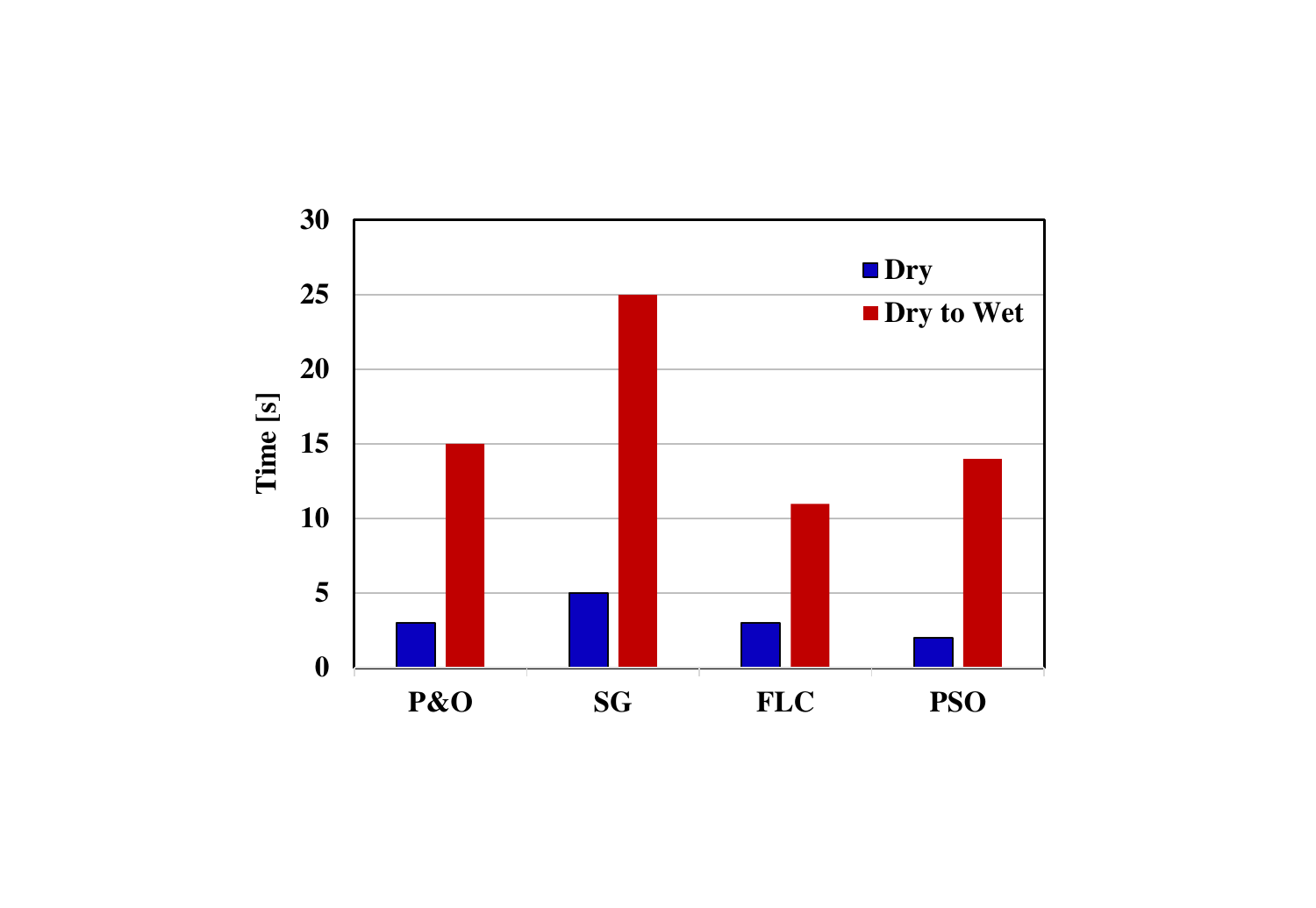}
        \label{fig:searching_time}
    }
         \subfigure[]
        {
        \includegraphics[width=0.45\textwidth,trim={4cm 3cm 3cm 4cm},clip]{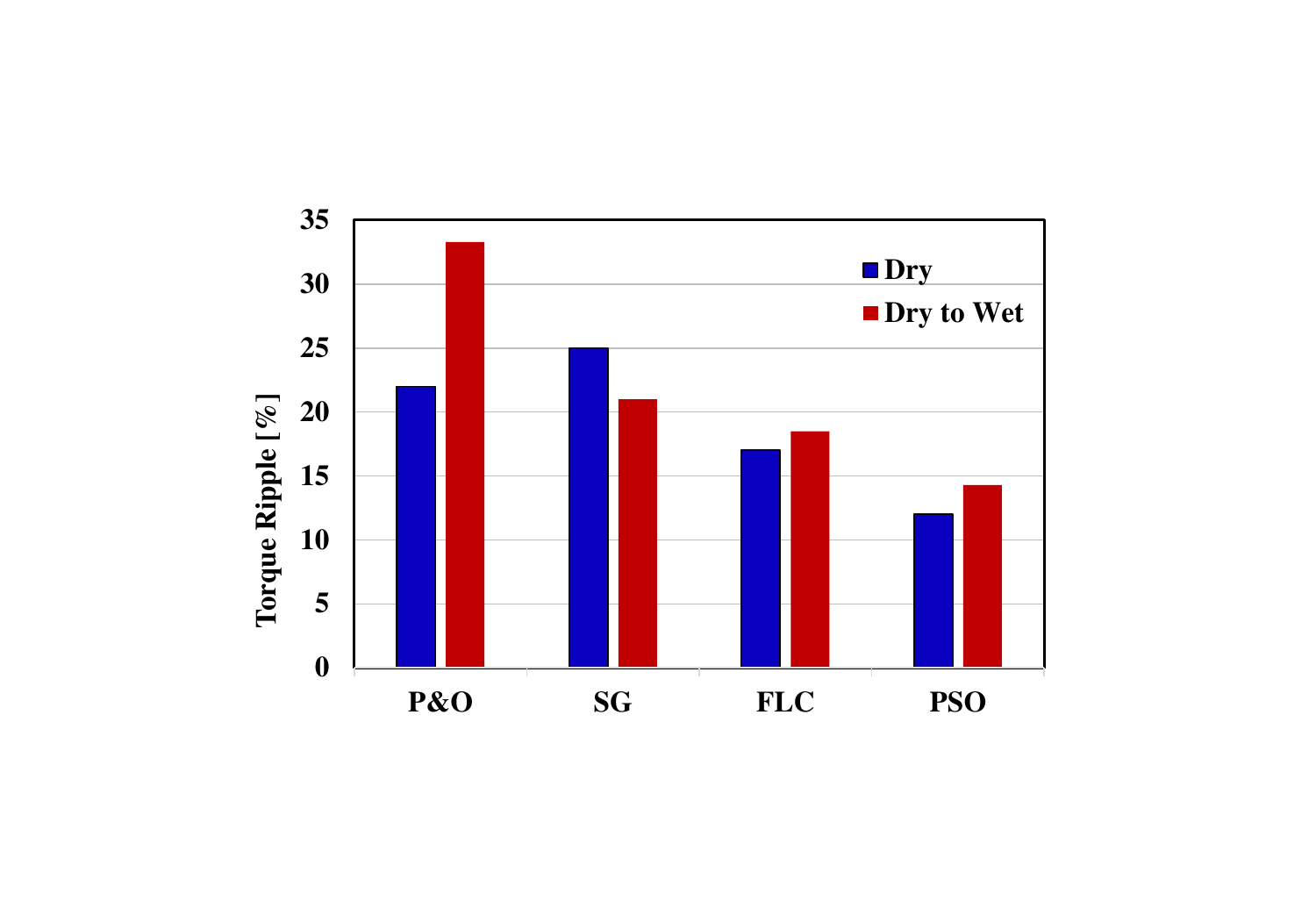}
        \label{fig:torque_ripples}
    }
        \caption{Experimental Results: Maximum Adhesion Tracking Methods (MAT) comparison for (a) Peak searching time; (b) Torque ripple.}
    \label{fig:comparison_charts}
\end{figure*}

Additionally, it is noted that MAT-PSO achieves completes the peak search in the dry condition in $t\approx 2$ s, which is the fastest response among all the methods. MAT-FLC achieves the maximum adhesion level faster ($t\approx 11$ s) in wet conditions, which is the fastest response among all the methods. MAT-SG is found to be the slowest peak searching method for both dry and wet tests with  $t\approx$ 5 s, and $t\approx 25$ s respectively (see Fig. \ref{fig:searching_time}).

The produced wheel motor torque ripples have been evaluated as it is a crucial issue in traction applications as it produces unwanted stress on the traction chain. As expected, MAT-P\&O and MAT-SG methods have high ripples due to the added perturbation during peak searching. MAT-PSO has a 5\% lower torque ripple compared to MAT-FLC in both dry and wet tests.    

\section{Conclusion}   \label{sec:conclusions}

Operating at maximum adhesion coefficient in railway traction is not straightforward due to the non-linear and unpredictable nature of wheel-rail contact.
  
Two new Maximum Adhesion Tracking (MAT) methods using Fuzzy Logic Control (MAT-FLC) and Particle Swarm Optimization (MAT-PSO) have been proposed in this paper. The main contributions of these methods are: 1) they don't require any adhesion coefficient estimation like classical ones and depend only on the available drive torque and slip velocity signals; 2) reduce the implementation complexity and avoid the parameter dependency of the adhesion estimators; 3) introduce and validate experimentally for the first time an artificial-intelligence-based algorithm (PSO) for maximum adhesion searching in railways.

Existing and proposed slip velocity control strategies have been simulated and validated experimentally under identical operating conditions where the wheel-rail contact point has been emulated using a scaled roller rig. 

It has been shown that the proposed methods have superior performance regarding the required searching time (60\% less), steady-state response, and torque ripples (20\% lower) compared to the classical MAT methods. In addition, the proposed MAT-FLC and MAT-PSO methods provide similar performance finding the maximum adhesion point under dry and wet wheel-roller conditions with stable slip velocity value at steady-state, the implementation of MAT-FLC being simpler.   


\bibliographystyle{IEEEtran}

\bibliography{Ref}

\end{document}